\documentclass{SciPost}
\hypersetup{
    colorlinks,
    linkcolor={red!50!black},
    citecolor={blue!50!black},
    urlcolor={blue!80!black}
}
\usepackage[bitstream-charter]{mathdesign}
\DeclareSymbolFont{usualmathcal}{OMS}{cmsy}{m}{n}
\DeclareSymbolFontAlphabet{\mathcal}{usualmathcal}
\fancypagestyle{SPstyle}{
\fancyhf{}

\fancyfoot[C]{\textbf{\thepage}}
}
\usepackage{algorithm}
\usepackage{algpseudocode}
\usepackage{setspace}
\usepackage{placeins}
\begin{document}
\pagestyle{SPstyle}

\begin{center}{\Large \textbf{\color{scipostdeepblue}{
Calo-VQ: Vector-Quantized Two-Stage Generative Model in Calorimeter Simulation\\
}}}\end{center}

\begin{center}\textbf{
Qibin Liu\textsuperscript{1 2$\star$},
Chase Shimmin\textsuperscript{3},
Xiulong Liu\textsuperscript{2},
Eli Shlizerman\textsuperscript{2}, 
Shu Li\textsuperscript{1} and
Shih-Chieh Hsu\textsuperscript{2$\dagger$}
%%%%%%%%%% END TODO: AUTHORS
}\end{center}
\begin{center}
{\bf 1} Tsung-Dao Lee Institute, Shanghai Jiao Tong University
\\
{\bf 2} University of Washington, Seattle
\\
{\bf 3} Yale University
\\[\baselineskip]
$\star$ \href{mailto:qibin.liu@sjtu.edu.cn}{\small qibin.liu@sjtu.edu.cn}\,,\quad
$\dagger$ \href{mailto:schsu@uw.edu}{\small schsu@uw.edu}
\end{center}

\section*{\color{scipostdeepblue}{Abstract}}
\boldmath\textbf{%
We introduce a novel machine learning method developed for the fast simulation of calorimeter detector response, adapting vector-quantized variational autoencoder (VQ-VAE). Our model adopts a two-stage generation strategy: initially compressing geometry-aware calorimeter data into a discrete latent space, followed by the application of a sequence model to learn and generate the latent tokens. Extensive experimentation on the Calo-challenge dataset underscores the efficiency of our approach, showcasing a remarkable improvement in the generation speed compared with conventional method by a factor of 2000. Remarkably, our model achieves the generation of calorimeter showers within milliseconds. Furthermore, comprehensive quantitative evaluations across various metrics are performed to validate physics performance of generation.
}

\vspace{\baselineskip}
\vspace{10pt}
\noindent\rule{\textwidth}{1pt}
\tableofcontents
\noindent\rule{\textwidth}{1pt}
\vspace{10pt}

\section{Introduction}
\label{sec:intro}
Simulation of detector response plays a crucial role for modern high energy physics experiments~\cite{foundation2018hep,software}. In the chain of physics event generation, from theory model to the realistic detector output, the detector simulation stands as the final and pivotal step for the interpretation of experimental observation and hunting for the un-expected signature. Notably, calorimeter, often the largest and most complex sub-system within the full detector, poses unique challenges for the simulation due to their extensive spatial scale, intricate structural composition, and the necessity for precise energy deposition.

Standard way simulating the calorimeter response entails accurate tracing of the propagation and evolution of secondary particles alongside their interactions with detector materials and finally the recording of energy deposited in each unit. Widely employed software like \textsc{GEANT4}~\cite{AGOSTINELLI2003250,ALLISON2016186,1610988} specializes in the simulation with Markov chain Monte Carlo process and is highly optimized in the computing and precision. However, as modern experiments continue to push the boundaries with increasingly large and high-granularity calorimeter designs, the computational demands of accurate simulation have emerged as a significant bottleneck~\cite{CERN-LHCC-2020-015}.

The rapid development of machine learning enlightened new hope offering novel avenues for fast simulation of calorimeter~\cite{HASHEMI2024100092}. Various methodologies have emerged, leveraging generative neural networks based on architectures like GAN~\cite{Paganini_2018,Buhmann_2021,ATL-SOFT-PUB-2020-006,Hashemi_2024}, normalizing flow~\cite{Krause_2023,krause2023caloflow,buckley2023inductive,ernst2023normalizing}, and diffusion~\cite{Mikuni_2022,amram2023denoising}.

In this paper, we introduce a novel generative model based on Vector-Quantized Variational Autoencoders (VQ-VAE)~\cite{VQ,birk2024omnijet,heinrich2024masked} to simulate calorimeter response, achieving a remarkable speedup exceeding 2000 times compared to \textsc{GEANT4}-based simulations on Calo-challenge datasets~\cite{michele_faucci_giannelli_2023_8099322,faucci_giannelli_2022_6366271,faucci_giannelli_2022_6366324}. Our model demonstrates realistic performance across physics variables\cite{Paganini_2018} such as total energy deposition and shower shape. Furthermore, we showcase the adaptability of the VQ-VAE approach across a variety of encoder-decoder architectures, ranging from fully-connected to convolutional networks. The former is optimal for smaller and irregular geometries, whereas the latter excels in handling datasets with regular structures and high granularity. Innovative designs including \textsc{FFT}-resampling (Fast Fourier Transform based resampling), cylindrical convolution, and physics-motivated normalization are developed to enhance performance on physics metrics.

The paper is structured as following: Section \ref{sec:dataset} outlines the features of the datasets, followed by Section \ref{sec:model} which summarizes the principle of two-stage design for calorimeter simulation, based on VQ-VAE architecture. Section \ref{sec:ae} and \ref{sec:gpt} detail the implementations of the two stages on different geometries. Section \ref{sec:eval} presents evaluation results encompassing both physics and computational aspects, with concluding remarks and future outlook discussed in Section \ref{sec:conclusion}.

%%%%%%%%%%%%%%%%%%%%%%%%%%%%%%%%%%%%%%%%%%%%%%%%%%%%%%%%%%%%%%%%%%%%%%%%%%%%%%%%%%%%%%%%%%%%%%%%%%%%%%%%%%%%%%%
\section{Calo-challenge Datasets}
\label{sec:dataset}
The datasets of Fast Calorimeter Simulation Challenge 2022 project~\cite{calocha}are used in this study. The datasets consists of 4 types of simulation from different incident particle and detector setup.
The different calorimeters encompassa range of voxel counts, varying from 368 to 40,500 which reflects various detector segmentation strategies. Specifically, one dataset captures the response of the calorimeter from photons, another from pions, and two others from electrons.

These datasets are generated with \textsc{GEANT4}, a Monte Carlo simulation program commonly used to model the particle-material interaction in the research of high energy physics. Dataset1-photon(referred to as "ds1\_photon" hereafter) and Dataset1-pion("ds1\_pion") originate from open datasets of calorimeter from ATLAS experiment. The showers induced by single photon or charged pion with 15 discrete incident energies ranging from 256 MeV to 4 TeV (in powers of 2) are recorded in the dataset. The energy deposited in calorimeter is voxelized based on the irregular detector geometry resulting in 368 voxels (5 layers) for photon dataset and 533 (7 layers) for pion. 

Dataset2("ds2") and Dataset3("ds3") capture the calorimeter response with higher granularity. The showers stem from single electron with incident energy logarithmically distributed from 1 GeV to 1 TeV. The detector's cylindrical geometry is described in three dimensions: depth ($Z$-direction where layers arranged), angular ($A$-direction) and radical($R$-direction). Dataset2 corresponds to 45*16*9=6,480 voxels while Dataset3 comprises 45*50*18=40,500 voxels. 

\begin{table}[]
\centering
\begin{tabular}{lrrrrr}
\hline 
 Dataset     & Layers & Radial Seg.         & Angular Seg.     & Voxels   & Condition Energy \\
 \hline
ds1\_photon  & 5      & irregular           &  irregular       &   368    & 15 bins      \\
ds1\_pion    & 7      & irregular           &  irregular       &   533    & 15 bins      \\
ds2          & 45     & 16                  &   9              &   6480   & continuous \\
ds3          & 45     & 50                  &  18              &   40500  & continuous \\
\hline
\end{tabular}
\caption{Details of four datasets from the Fast Calorimeter Simulation Challenge 2022.}
\label{tab:dataset}
\end{table}

The configuration and statistical characteristics of each dataset are detailed in Table~\ref{tab:dataset}. Each of the 4 types of datasets is given in the format of several statistically independent and equally divided files which are naturally used either for the training of generative model or the assessment of quality of generated data. For all the study in the following, half of the statistics are used for training and half for the reference of performance evaluation, of each type of dataset.

%%%%%%%%%%%%%%%%%%%%%%%%%%%%%%%%%%%%%%%%%%%%%%%%%%%%%%%%%%%%%%%%%%%%%%%%%%%%%%%%%%%%%%%%%%%%%%%%%%%%%%%%%%%%%%%
\section{Two-stage Modeling Approach}
\label{sec:model}

\begin{figure}[]
    \centering
    \includegraphics[width=.8\linewidth]{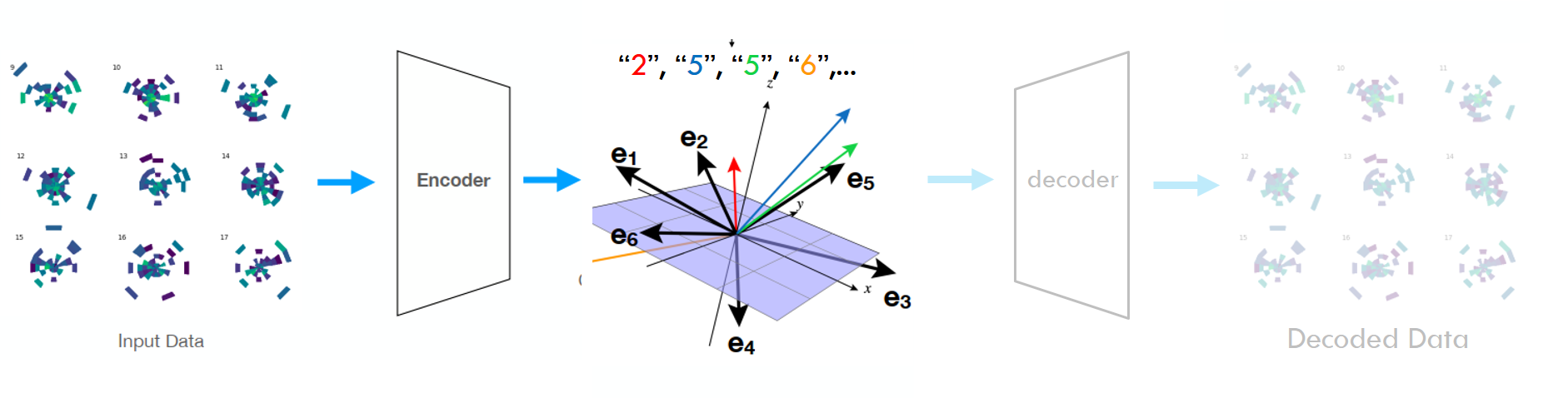}
    \includegraphics[width=.8\linewidth]{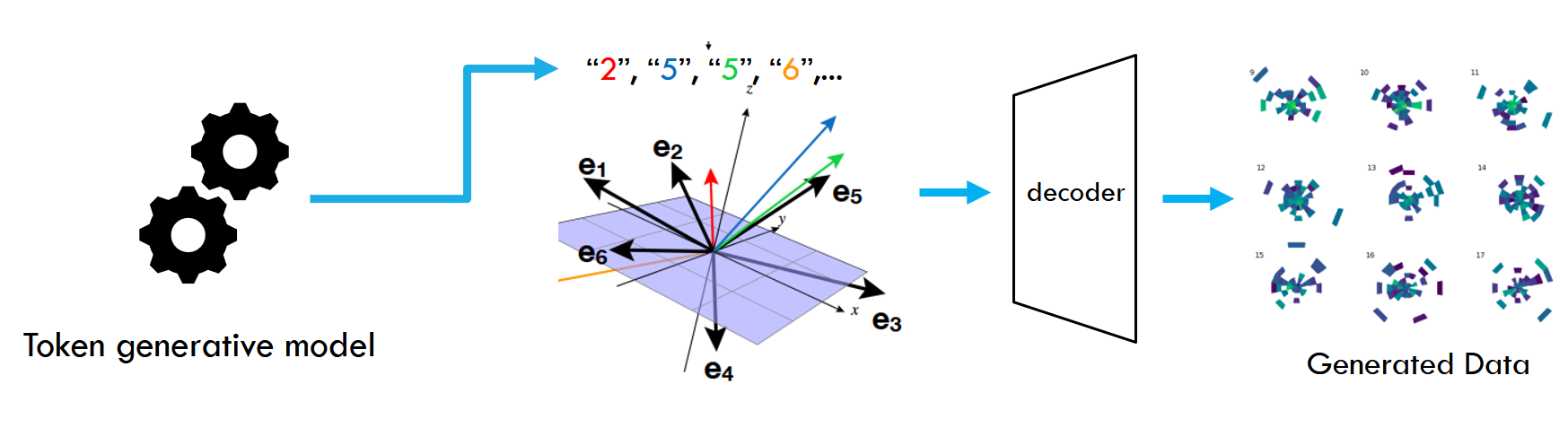}
    \caption{Demonstration of the vector-quantization based two-stage generative model. The upper and lower parts show the two stages of the model, respectively.}
    \label{fig:caloVQ_demo}
\end{figure}

The calorimeter detector with high granularity typically entails large number of channels (voxel/pixel) of the response up to tens of thousand. Direct sampling in such high dimensionality and highly sparse space is frequently inefficient and impractical. To address this challenge, we propose a two-stage method, as depicted in Figure~\ref{fig:caloVQ_demo}, drawing inspiration from advancements in the field of computer vision \cite{VQGAN}. 

The first stage is designed to reduce the dimensionality of data. The encoder is utilized to transform the input into a different representation in latent space followed by the decoder which reconstruct the input. In order to achieve higher compression and more effective usage of latent space, vector quantization technique is implemented in the latent space as in the model of Vector Quantized Variational Auto Encoder (VQ-VAE). The same architecture had been shown to be efficient in the jet physics for the tokenization of physics object and benefits the classification and generation task \cite{birk2024omnijet,heinrich2024masked}. In our study we focus on the generation of raw detector hit with higher dimensional and low-level information. And we found that the encoder loss is not efficient to describe the distance between input and reconstructed data in the sense of high resolution calorimeter response. The discriminator loss is added to improve the quality of reconstruction from VQ-VAE. 

This stage enables the learning and sampling of the probabilistic distribution in a highly reduced and regularized space. While the second stage focuses on the modeling of latent space which is a sequence of discrete codes with fixed length. GPT-based model is adapted to sample the latent codes conditioned on the incident energy. As discussed in the previous section, the normalization factor, which controls the energy deposition of each calorimeter layer, is digitized into discrete codes and sampled at the same time of latent codes. 

\section{First Stage: Compression with VQ-VAE}
\label{sec:ae}
The VQ-VAE model is tailored for utilization in the first stage. Following the standard notation, the input data, energy deposited in each individual voxel $i$, is named $E_i$ and after pre-processing $x_i$. The output of VQ-VAE denoted as $\tilde{x_i}$. The index $i$ runs over all the numbers as shown in the last column of Table~\ref{tab:dataset}. The global condition incident energy, is denoted as $E_c$. 

The first stage model includes the encoder network $q(z|x;E_c)$ which parameterize the posterior distribution of latent variable $z$ given the input $x$, decoder network $p(x'|z;E_c)$ which runs inversely and the prior distribution $P(z)$ which will be modelled by the second stage.

In contrast to the Variational Auto Encoder (VAE)~\cite{VAE}, the latent distribution $P(z)$ is not re-parameterized by independent Gaussian but discrete index of quantized embedding vector. The embedding vector is defined in the space $e \in R^{D}$ where $D$ is the size of latent dimension. The output of encoder of each input is quantized to $L$ learnable vectors where $L$ denotes the width of encoder output. In total $K$ unique vectors are learned and they make up a "codebook", relate the index $k$ of vector to the actual latent vector $e_k$. The index is also termed "code" hereafter. The latent space is then represented as,

\begin{equation}
z \equiv \{e_{k(i)}\}~\text{where}~
i=1,2,...,L~\text{and}~
k(i) \in [1,K]
\label{eq:vq}
\end{equation}

The back-propagation of the non-differentiable quantization operation is implemented simply copies the gradients from the decoder to the encoder~\cite{VQ} described as

\begin{equation}
L_{VQ} = ||x-p(q(x))||^2 + ||sg[q(z|x)] - e_k||^2 + ||q(z|x) - sg[e_k]||^2
\label{eq:vq}
\end{equation}

where $sg[X]$ denotes the stop-gradient operator which will not take the gradient of $X$ into calculation. The first term is named "reconstruction loss" which is designed to minimize the distance between input and output of the whole VQ. The second term denotes the quantization loss which optimize the codebook to better represent the latent space. The last term is called commitment loss which limits the arbitrarily growth of embedding space and makes the encoder output commit to a sets of vectors.

It is found in the study that the VQ-VAE model trained only with loss as described in Formula~\ref{eq:vq} is not sufficient to obtain high fidelity due to the limitation of L1/L2 loss. This is also observed in~\cite{VQGAN}. So an additional discriminative regularization (loss)~\cite{lamb2016discriminative,mentzer2020highfidelity} is implemented and updated during the training as

\begin{equation}
L_{GAN}=log D(x) + log(1-D(p(q(x))))
\label{eq:gan}
\end{equation}

where the $D$ denotes the discriminative network which is trained simultaneously as the VQ network to capture the subtle difference between the decoded output and input.

The total training objective of the first stage model is

\begin{equation}
\underset{q,p,\mathcal{Z}}{arg\,min}\underset{D}{max}[L_{VQ}(q,p,\mathcal{Z})+\lambda L_{GAN}(D)]
\label{eq:minmax}
\end{equation}

where the $q,p$ denote the encoder and decoder, $\mathcal{Z}$ denotes the trainable codebook and $\lambda$ denotes the adaptive factor~\cite{VQGAN} to balance the updating of VQ and the discriminator. 

\subsection{Data Pre-processing} % cyc conv and FFT resampling
\label{ssec:pre}
The data is pre-processed before feeding into the first stage model. The data is firstly normalized per shower with the normalization factor denoted as $R$. The factor is calculated per shower and needs to be sampled if new shower being generated, which will be discussed in the next section of the second stage~\ref{sec:gpt}. In the first stage the $R$ factor is assumed known information. Then the data is transformed logarithmically with proper linear scaling and shifting. 

The pre-processed is described as formula,

\begin{equation}
x_i=\frac{1}{c}\log\left(a + b \frac{E_{i}}{R_{l(i)}}\right)
\label{eq:pre}
\end{equation}

where $i$ is the index of voxel and $E_i$ is the original value of each voxel. $a,b,c$ are hyper-parameter tuned according to inputs and the value documented in Appendix~\ref{asec:arch}. The $R_{l(i)}$ denotes the corresponding normalization factor for each voxel location with different normalization scheme. 

Two normalization methods are tested. Global method normalizes the whole detector energy to 1 so that $R_{l(i)}=R=\sum_{j}{E_j}$ where $j$ runs over all the voxels. Layer-wise method however performs normalization per detector layer and $R_{l(i)}=\sum_{l(j)=l(i)}{E_j}$ where the summation collects all the voxel indices in the same layer $l(i)$. The layer-wise normalization takes the fact into consideration that the energy deposition is highly different from layer to layer of the detector, sometime over order of magnitude in one shower. Utilization of multiple normalization factors can achieve better performance with the cost of more need to be learned and sampled in the the second stage. This method basically trades speed for quality.

The input after pre-processing implies $x_i \simeq log(E_i/R_{l(i)})$ and $\sum e^{x_i^*} = 1$ where $x^*$ relates to $x$ with only scaling and shifting using constants $a,b,c$. This inspires us using the softmax~\cite{bridle1990probabilistic} activation in the last layer of decoder. It is validated in experiments that the softmax makes the training more stable and well performed than other activation. 

The post-processing defines the inverse operation and described as 
\begin{equation}
\tilde{E_i}=\tilde{x}_i \tilde{R}_{l(i)}
\label{eq:pre_2}
\end{equation}

where the $\tilde{x}_i$ denotes the decoder output and $\tilde{R}_{l(i)}$ the corresponding normalization factor which is sampled in second stage model.

\subsection{Convolutional Network for Cylindrical Geometry} 
\label{ssec:cyc}
\begin{figure}[]
    \centering
    \includegraphics[width=.6\linewidth]{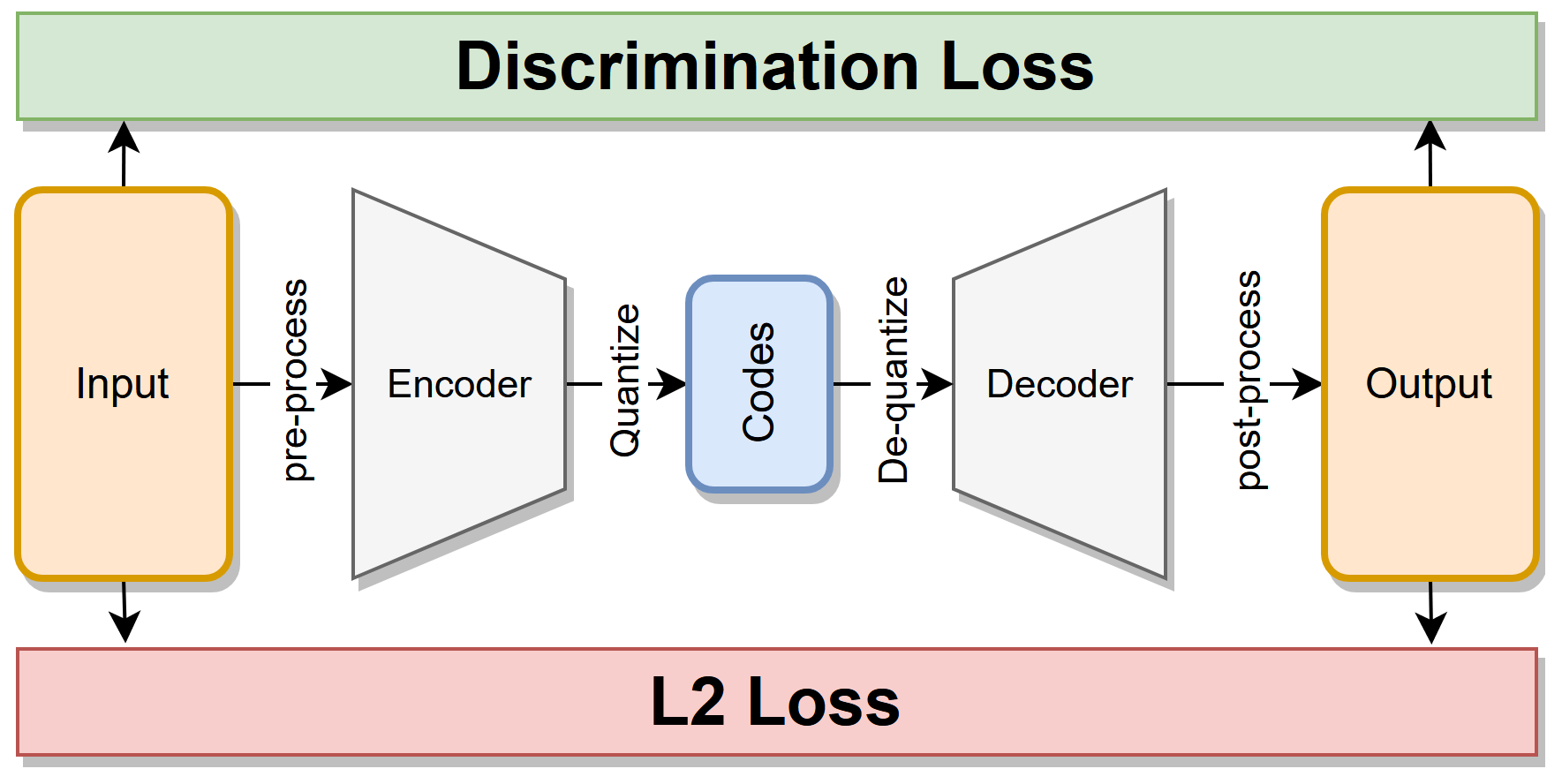}
    \caption{Architecture of Stage-1 Model.}
    \label{fig:S1_arch}
\end{figure}

To enhance performance on particular calorimeter geometries, the encoder and decoder model are implemented with various layers on different datasets.
For dataset1(photon) and dataset1(pion), 1D convolutional layer and fully connected layer are combined to handle the irregular geometry. The architecture of detaset1 model is shown in the Figure~\ref{fig:S1_arch}

\begin{algorithm}
\caption{Cylindrical 2D Convolution}
\label{alg:cconv}
\setstretch{1.25}
\begin{algorithmic}
\Require
\State $x \in \mathbb{R}^{C_{in} 
\times Z \times A}$
\Comment{Input data as 2D array in $C_{in}$ channels}

\State $W \in \mathbb{R}^{C_{out} \times C_{in} \times k_z \times k_\alpha}$ \Comment{Network parameter for kernels}

\Ensure $y = \mathrm{\texttt{cyl\_conv2d}}(x, W)$
\State $p_\alpha \gets \lfloor \frac{(k_\alpha-1)}{2} \rfloor$ \Comment{Compute padding for angular dimension}
\State $a_{ij} \gets \left( x_{ijk} \right)_{k=1}^{p_\alpha-1}$
\State $b_{ij} \gets \left(x_{ijk}\right)_{k=A-p_\alpha+1}^{A}$
\State $y \gets \mathrm{\texttt{concat}}( [b, x, a], \mathrm{\texttt{axis}}=2)$ \Comment{Pad x with cyclic boundary} 
\State $y \gets \mathrm{\texttt{conv2d}}(y, W)$ \Comment{Normal convolution operation}
\end{algorithmic}
\end{algorithm}

\begin{algorithm}
\caption{Transposed Cylindrical 2D Convolution}
\label{alg:cconv-transpose}
\setstretch{1.25}
\begin{algorithmic}
\Require
\State $x \in \mathbb{R}^{C_{in} 
\times Z \times A}$
\Comment{Input data as 2D array in $C_{in}$ channels}

\State $W \in \mathbb{R}^{C_{in} \times C_{out} \times k_z \times k_\alpha}$ \Comment{Network parameter for kernels}

\Ensure $y = \mathrm{\texttt{cyl\_conv\_transpose2d}}(x, W)$
\State $y \gets \mathrm{\texttt{conv\_transpose2d}}(x, W)$ \Comment{Normal transpose-convolution operation}
\State $e_\alpha \gets \lfloor \frac{(k_\alpha-1)}{2} \rfloor$ \Comment{Compute excess for angular dimension}
% \State $e_\alpha \gets k_\alpha - s_\alpha$ \Comment{Compute excess for angular dimension}
% \State $a_{ij} \gets \{\}$
% \State $y \gets \mathrm{\texttt{concat}}( [a, x, b], axis=2)$ \Comment{Pad x with cyclic boundary} 
\State $y_{ij} \gets \left( y_{ijk} \right)_{k=e_\alpha+1}^{e_\alpha+A}$ \Comment{Trim the excess} 
% \State $y \gets \mathrm{\texttt{conv2d}}(y, W, (p_z, 0))$
\end{algorithmic}
\end{algorithm}

\FloatBarrier

For dataset2 and 3 the 2D convolutional layer is adopted. The 3D calorimeter data is interpreted as multi-channel 2D data considering that the transnational symmetry only exists in the $Z$- and $A$- direction, as described in Section~\ref{sec:dataset}. $R$- direction is treated as channel dimension and this image-like data is denoted "calo-image" in the following. It is notable that the translation in angular direction is periodical bounded. A special cylindrical convolution operation is defined on the $(Z,A)$ space, with vanilla translation in one direction and circular wrapped in another. The operation is described in Algorithm \ref{alg:cconv} and \ref{alg:cconv-transpose} where the handling of vanilla padding in the $Z$-direction is trivial and omitted.

\begin{figure}[]
    \centering
    \includegraphics[width=.6\linewidth]{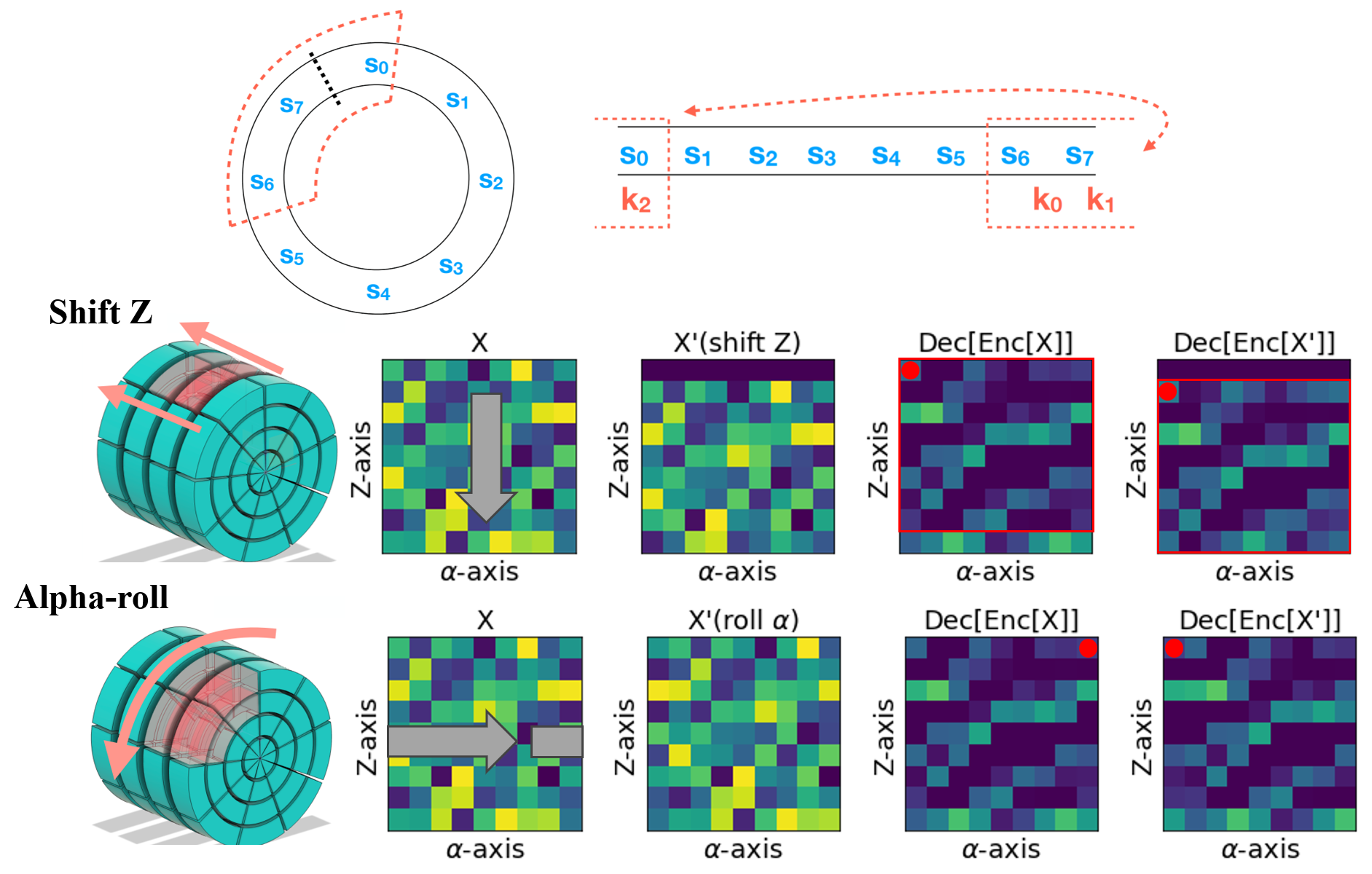}
    \caption{Validation of cylindrical convolution. Equivariant property is well kept for different transformation.}
    \label{fig:caloVQ_cyconv}
\end{figure}

All the layers for the dataset2/3 model operating on $(Z,A)$ space are implemented with cylindrical convolution. The equivariant property of the calorimeter data is well kept when mapped in the cylindrical coordinate, as shown in Figure~\ref{fig:caloVQ_cyconv}. 

\begin{figure}[]
    \centering
    \includegraphics[width=.6\linewidth]{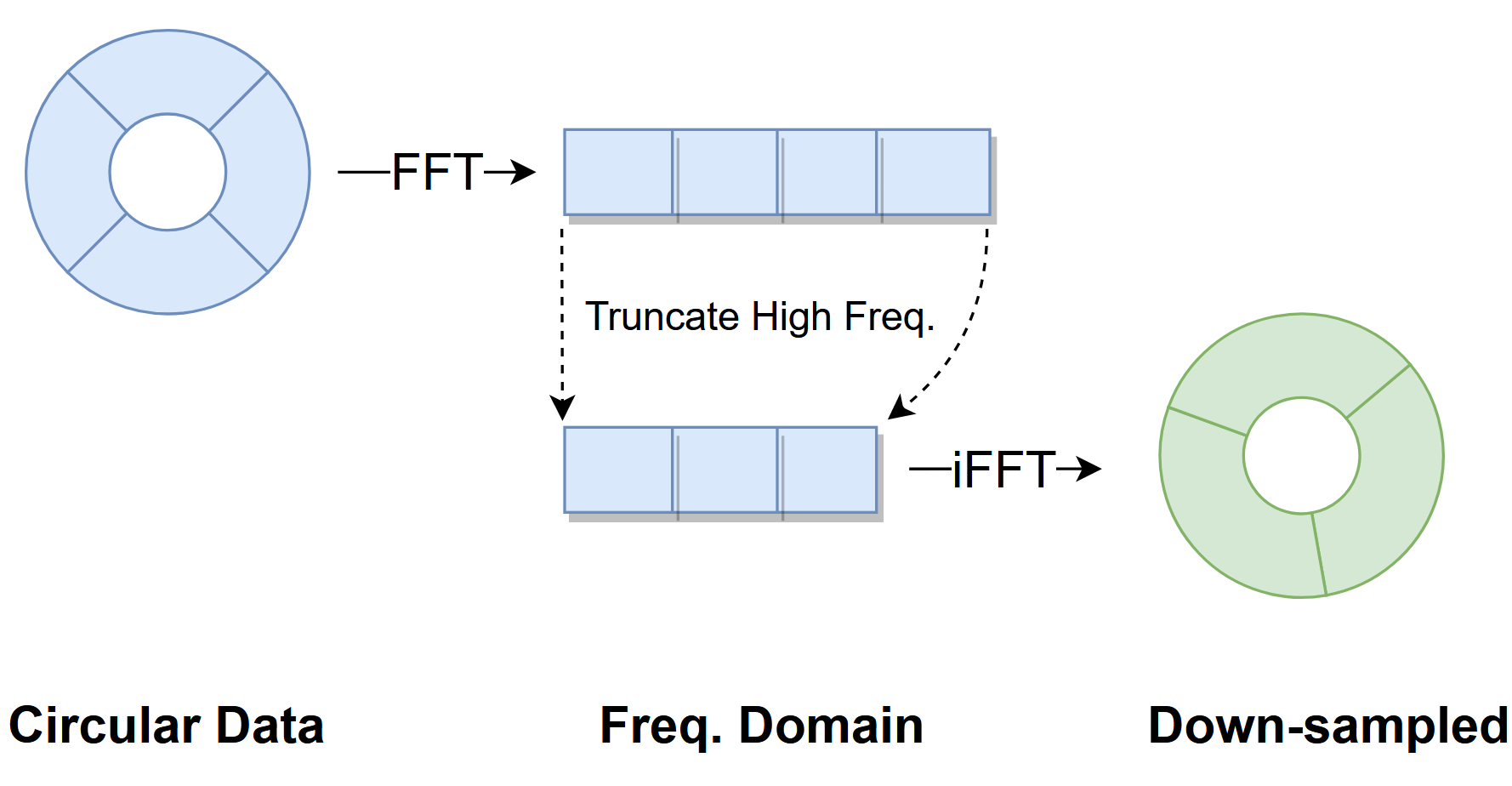}
    \caption{Demonstration of FFT down-sampling. }
    \label{fig:caloVQ_FFT}
\end{figure}

To achieve arbitrary up- or down-sampling along the angular direction with circular or periodic boundaries, we employ FFT-resampling, as illustrated in Figure~\ref{fig:caloVQ_FFT}. The data is firstly transformed into frequency space with discrete FFT along the angular direction, then resampled with targeted resolution and inversely transformed back.

\section{Second Stage: Latent Sampling with GPT} 
\label{sec:gpt}
In the second stage, we leverage auto-regressive model to learn and sample the prior distribution of the discrete latent space.

GPT~\cite{radford2019language,minGPT}, general pre-training transformer, is highly effective at modeling the distribution of discrete token space. It is widely utilized in natural language generation, as well as image and audio generation tasks. GPT employs multi-layer transformer decoder blocks with a causal mask to capture the sequence information of tokens. This makes it well-suited for the task of the second stage in the fast calorimeter simulation context, as it can efficiently sample a fixed length of discrete codes based on the given conditions.

One notable advantage of the token model is the flexibility to append or co-generate tokenized auxiliary information along with the token sequence. This facilitates natural establishment of mutual information exchange. In our model we explore to tokenize the continuous energy information and co-generate with the latent codes from the first stage. One example is the shower-wise normalization factor(s) $R$ which is mentioned in pre-processing section. The co-generation of $R$ with the latent codes improves the physics performance as it allows for better consideration of both the energy scale and spatial distribution.

The details of tokenization and application for specific dataset are described in Sec~\ref{ssec:steps} and Sec~\ref{ssec:pre}.

The complete sequence for each sample to learn in the second stage is composed like

\begin{equation}
\underbrace{A_1 A_2 ...}_{aux~tokens}\overbrace{B_1 B_2 B_3 B_4 ....}^{shower~codes}
\label{eq:codes}
\end{equation}

and there is no special spacer, separator or terminator inside.

\subsection{Training and Sampling of the Two-stage Model}
\label{ssec:steps}
The training of the two stages of calo-VQ model is performed sequentially, taking the advantage to optimize each stage with well-defined and decomposed tasks. The training of the first stage model as autoencoder-like architecture, focuses on achieving high-quality reconstruction. Meanwhile the training of the second stage model is executed in a sliding manner, aiming to maximize the probability of predicting the next correct token based on all preceding ones.

As mentioned, auxiliary information can be sampled together with the quantized codes in the second stage model. Here, we incorporate the $R$ factors as auxiliary information tokenized with the digitization. Through experimentation, we determined that utilizing 20bit (approximately 6 decimal digits) ensures high-fidelity sampling of energy. With a codebook size of 1024, each scalar value corresponds to an additional 2 tokens in the sequence to be learned which does not impose significant overhead. The detailed setup such as number $R$ codes for different model is listed in Table~\ref{tab:caloVQ_parameters}

%%%%%%%%%%%%%%%%%%%%%%%%%%%%%%%%%%%%%%%%%%%%%%%%%%%%%%%%%%%%%%%%%%%%%%%%%%%%%%%%%%%%%%%%%%%%%%%%%%%%%%%%%%%%%%%
\section{Evaluation}
\label{sec:eval}
The fast calorimeter simulation model undergoes evaluation by comparing the distribution of physics variables across a predefined set of events between generated data and testing data, conditioned with same incident particle and same energy range. The latter, produced with the same GEANT4 setup as the training dataset but never seen during training. Physics variables, including total energy, layer energy, shower shape variables and others are computed. The complete list of variable definitions is outlined in Table~\ref{tab:v}. Given the statistical nature of these physics variables, their distributions are histogrammed to mitigate fluctuation effects. The disparity between two histograms is quantified using the "Separation Metric"~\cite{Diefenbacher_2020},

\begin{equation}
Sep. \equiv \sum_{i=1 .. Nbin} \frac{(h'_i-h_i)^2}{2(h'_i+h_i)}
\label{eq:eva_1}
\end{equation}

$h'_i$ and $h_i$ denote the i-th bin content in the histogram of sampled and truth distribution, respectively. 

\begin{table}[]
\centering
\begin{tabular}{ccc}
\hline
Symbol & Name & Definition  \\
\hline
$E_i$                 & Cell energy for index $i$            &   (raw data)              \\
$E_{inc}$                 & Incident energy            &   (condition)              \\
$\eta_i (K)$/$\phi_i (K)$                & Location(mm) in $\eta$/$\phi$ at cell $i$ in layer $K$            &   (geometry)              \\
$\Delta \eta_i (K)$/$\Delta \phi_i (K)$          & Segment.(mm) in $\eta$/$\phi$ of cell $i$ in layer $K$            &   (geometry)              \\
$E_{tot}$                 & Total energy               &   $\sum E_i$       \\
$E_{tot}/E_{inc}$                 & Total energy response               &   $\sum E_i/E_{inc}$       \\
$E(K)$                     & Layer energy (in layer $K$)     &  $\sum_{i \in K} E_i$ \\
$\overline{\eta (K)}$  & Energy Center (EC) in $\eta$ (in layer $K$)   &  $\sum_{i \in K} \eta_i E_i / E(K)$   \\
$\sigma_{\eta (K)}$    & Shower Width in $\eta$ (in layer $K$)    &  $\sum_{i \in K} \Delta \eta_i E_i /E(K)$  \\
$\overline{\phi (K)}$  & Energy Center (EC) in $\phi$ (in layer $K$)   &  $\sum_{i \in K} \phi_i E_i / E(K)$   \\
$\sigma_{\phi (K)}$    & Shower Width in $\phi$ (in layer $K$)    &  $\sum_{i \in K} \Delta \phi_i E_i /E(K)$  \\
\hline
\end{tabular}
\caption{Symbol and definition of physics variables. }
\label{tab:v}
\end{table}

\subsection{Dataset 1: photon and pion}
\label{ssec:ds1}
\begin{figure}[]
    \centering
    \includegraphics[width=.8\linewidth]{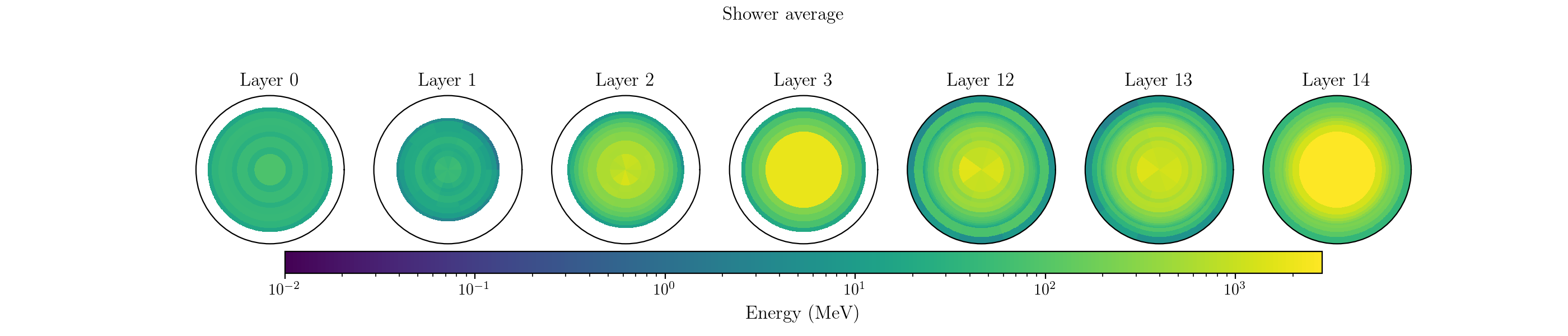}
    \includegraphics[width=.8\linewidth]{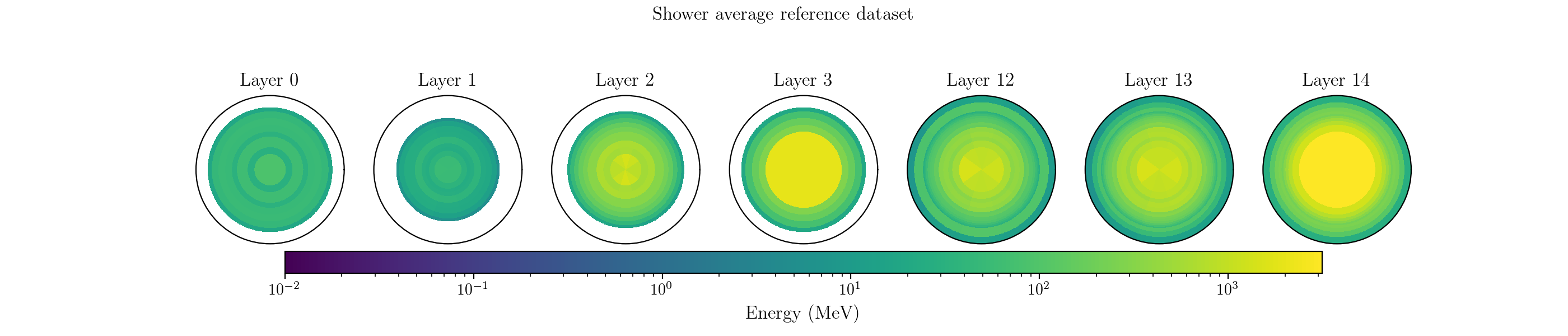}
    \caption{Average energy deposition of calorimeter cells. Each shower is induced by single incident of pion. Generated on the top and reference on the bottom. }
    \label{fig:ds1_pion_avg}
\end{figure}

\begin{figure}[]
    \centering
    \includegraphics[width=.3\linewidth]{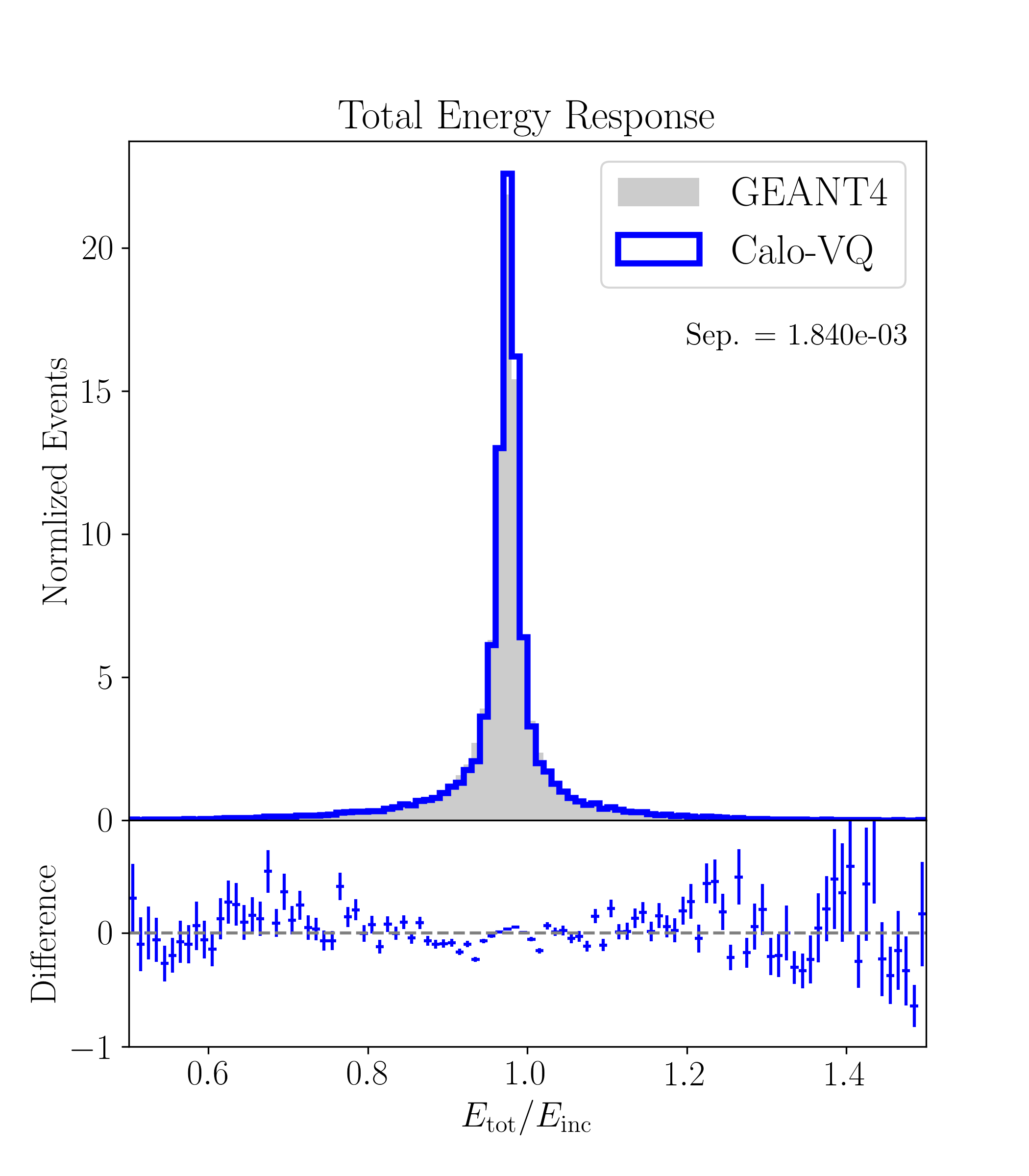}
    \includegraphics[width=.3\linewidth]{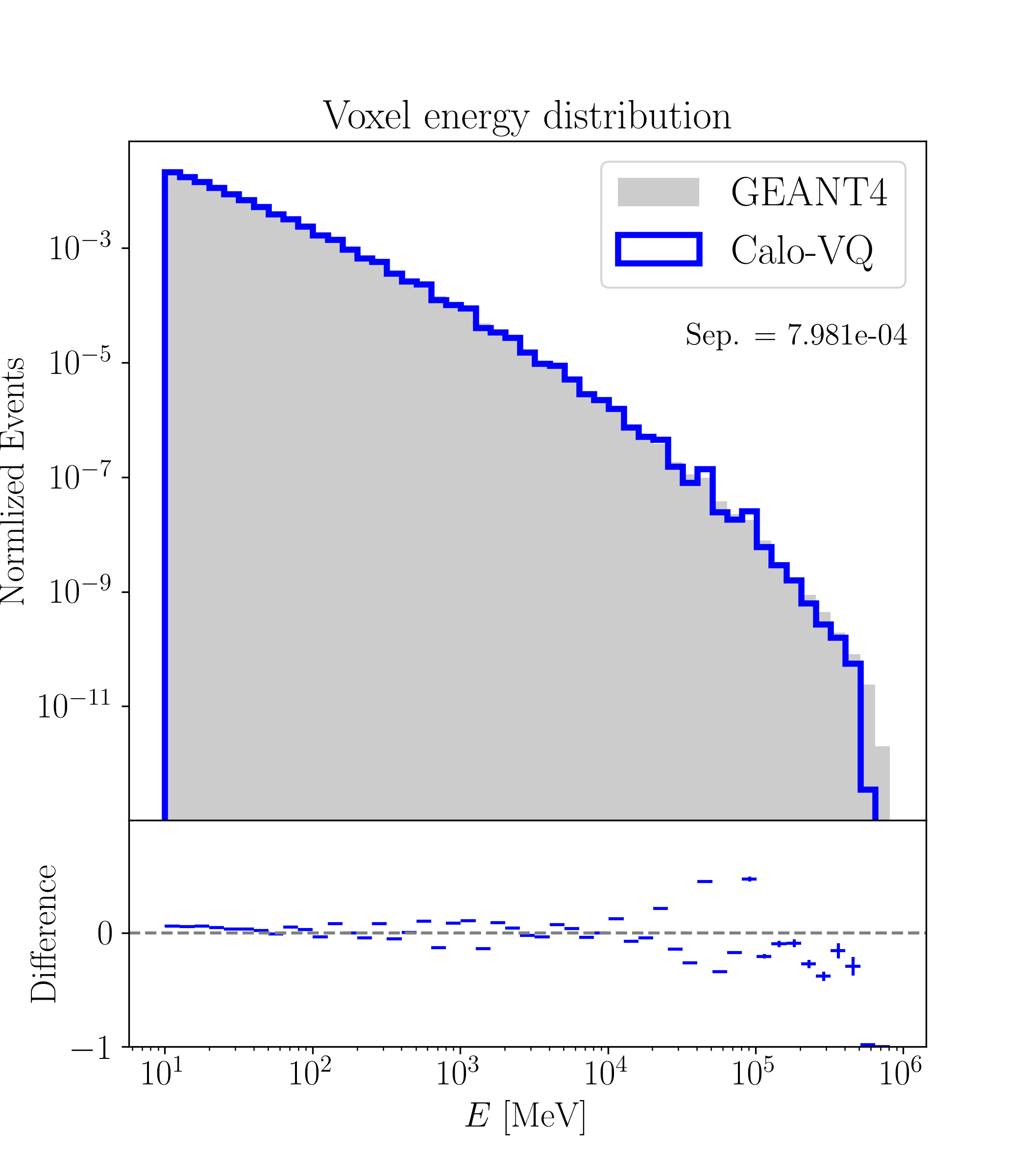} \\
    \includegraphics[width=.3\linewidth]{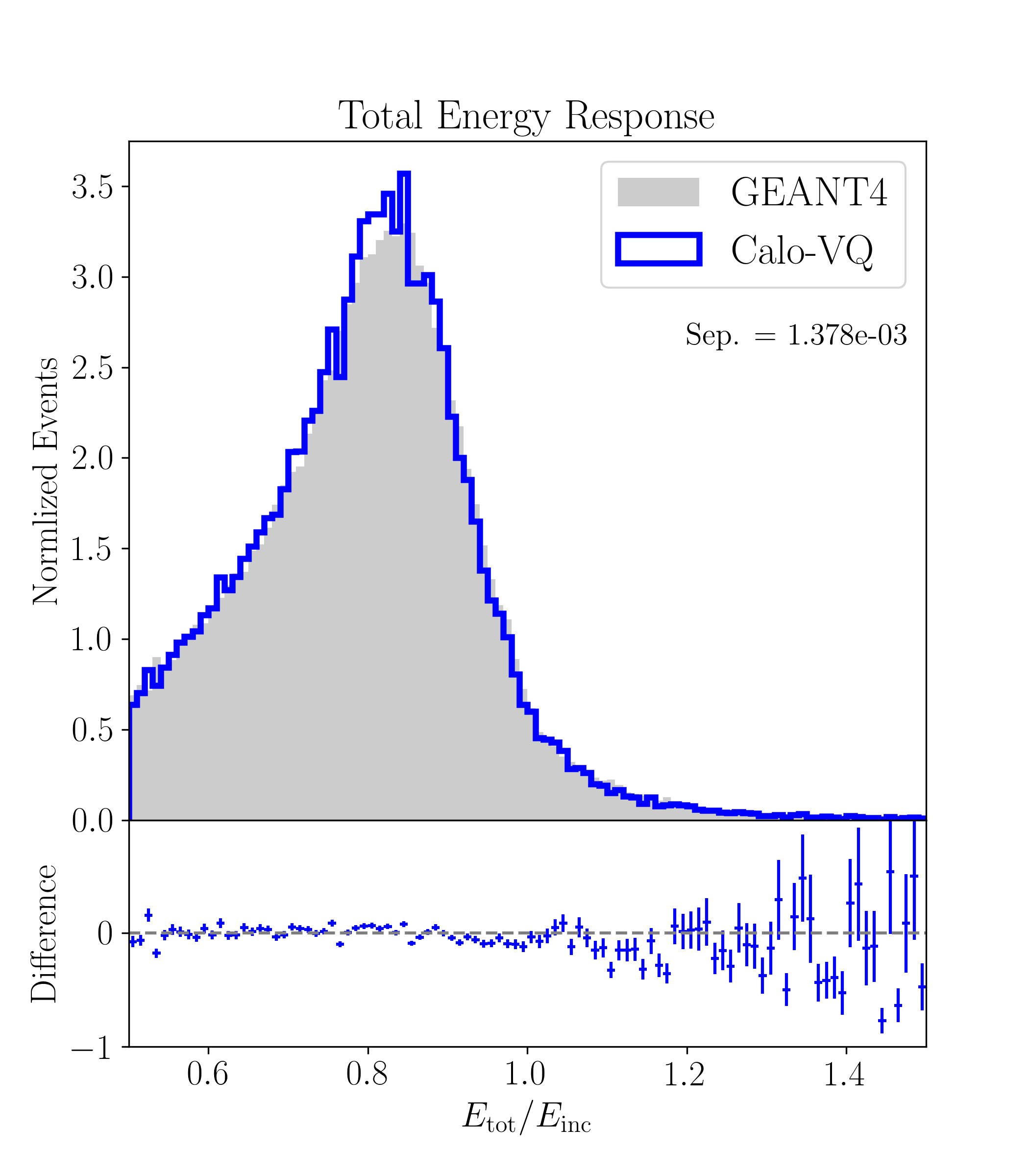}
    \includegraphics[width=.3\linewidth]{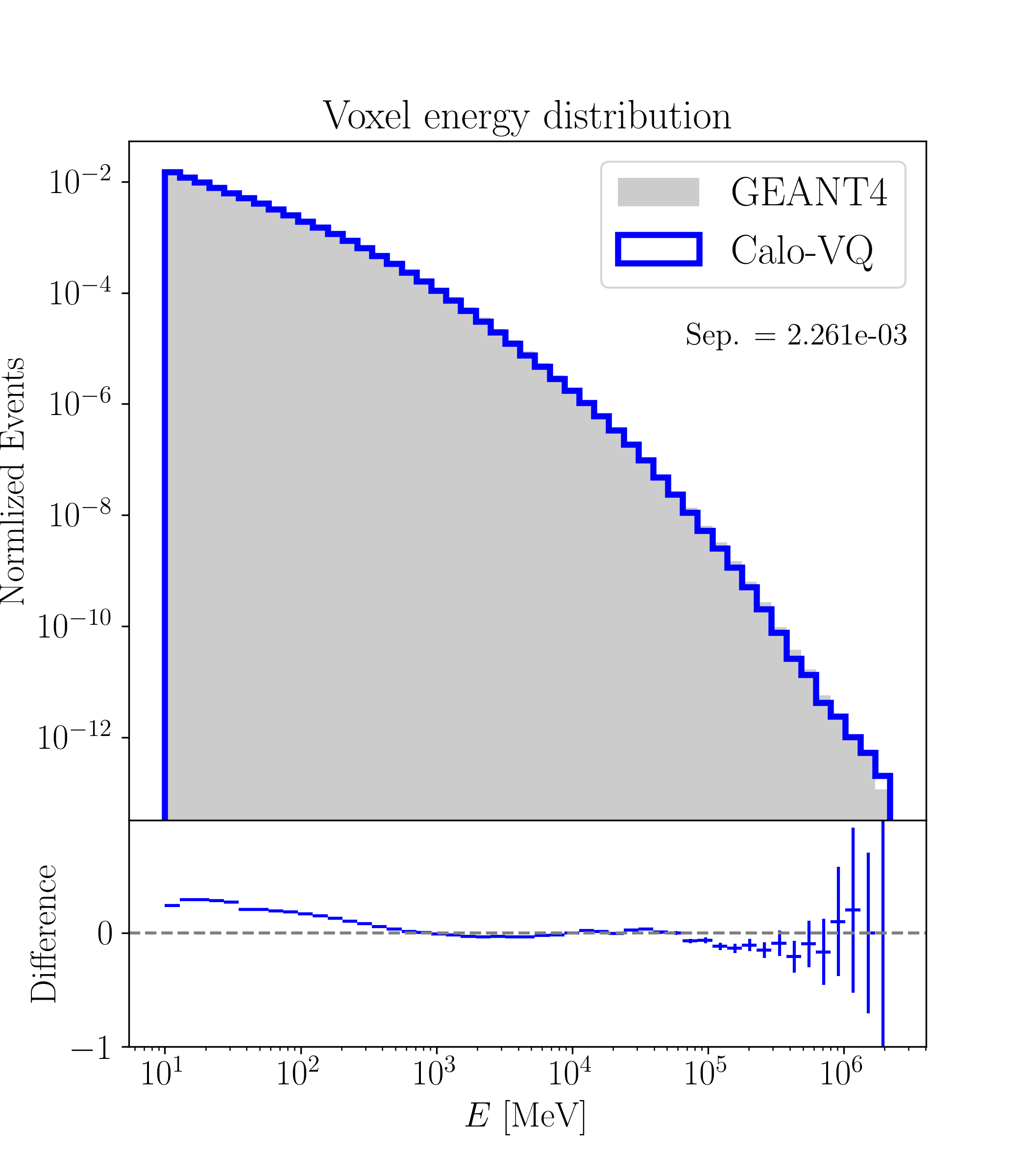}
    \caption{Distribution of cell energy and total energy response of photon(top) and pion(bottom) shower. Generated is shown in orange line and reference from GEANT4 shown in solid blue.}
    \label{fig:ds1_E}
\end{figure}

\begin{figure}[]
    \centering
    \includegraphics[width=.3\linewidth]{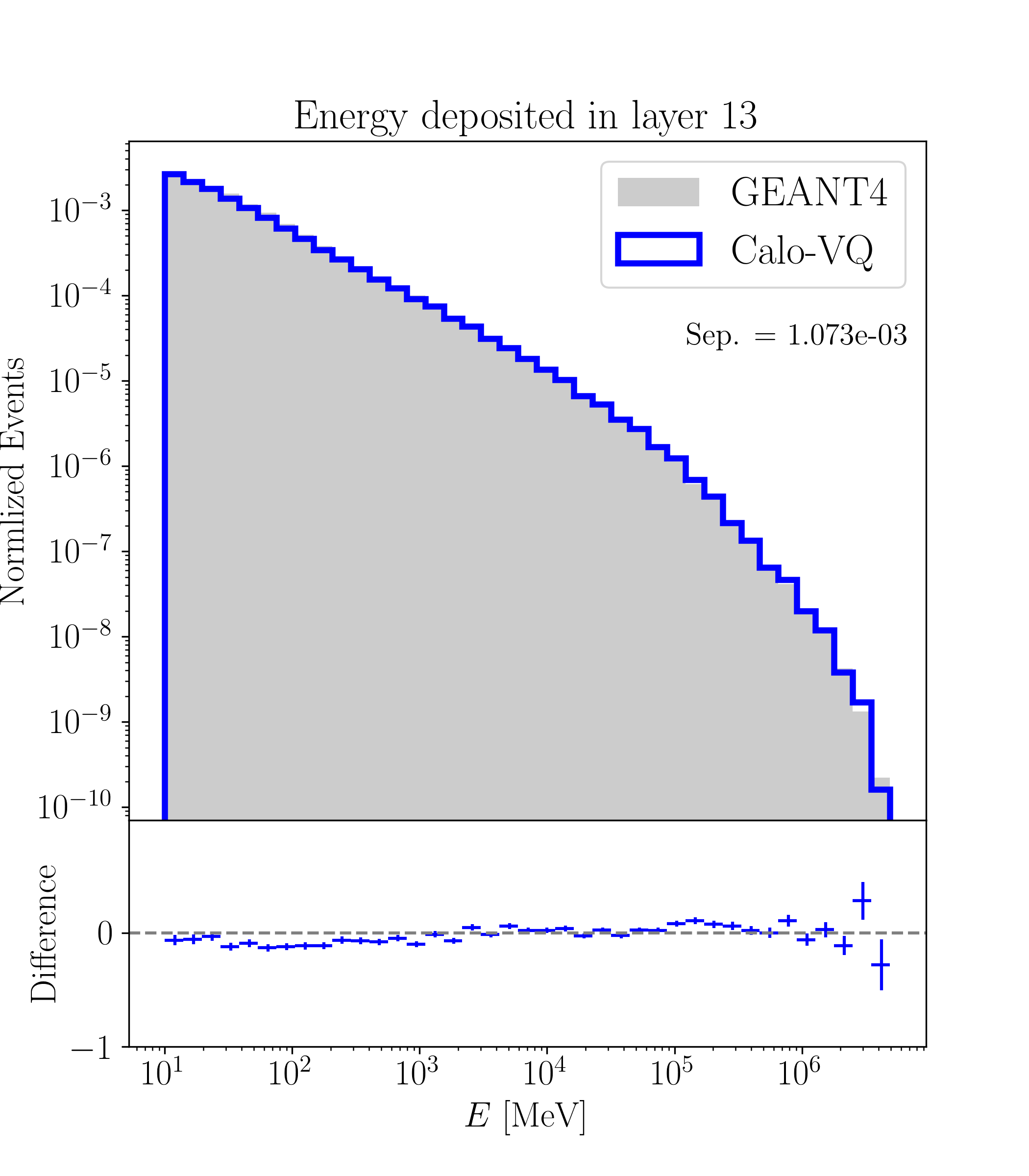}
    \includegraphics[width=.3\linewidth]{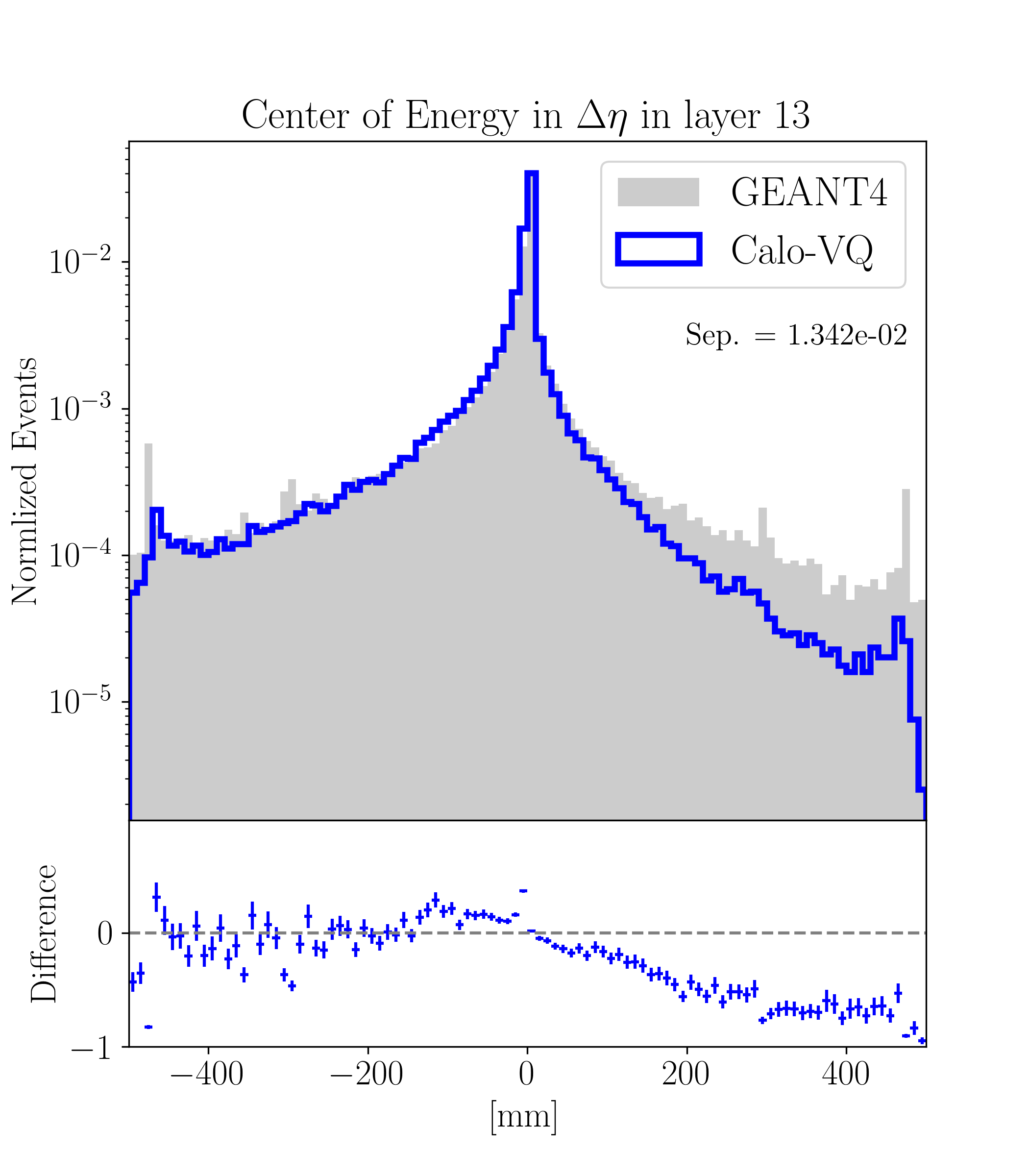}
    \includegraphics[width=.3\linewidth]{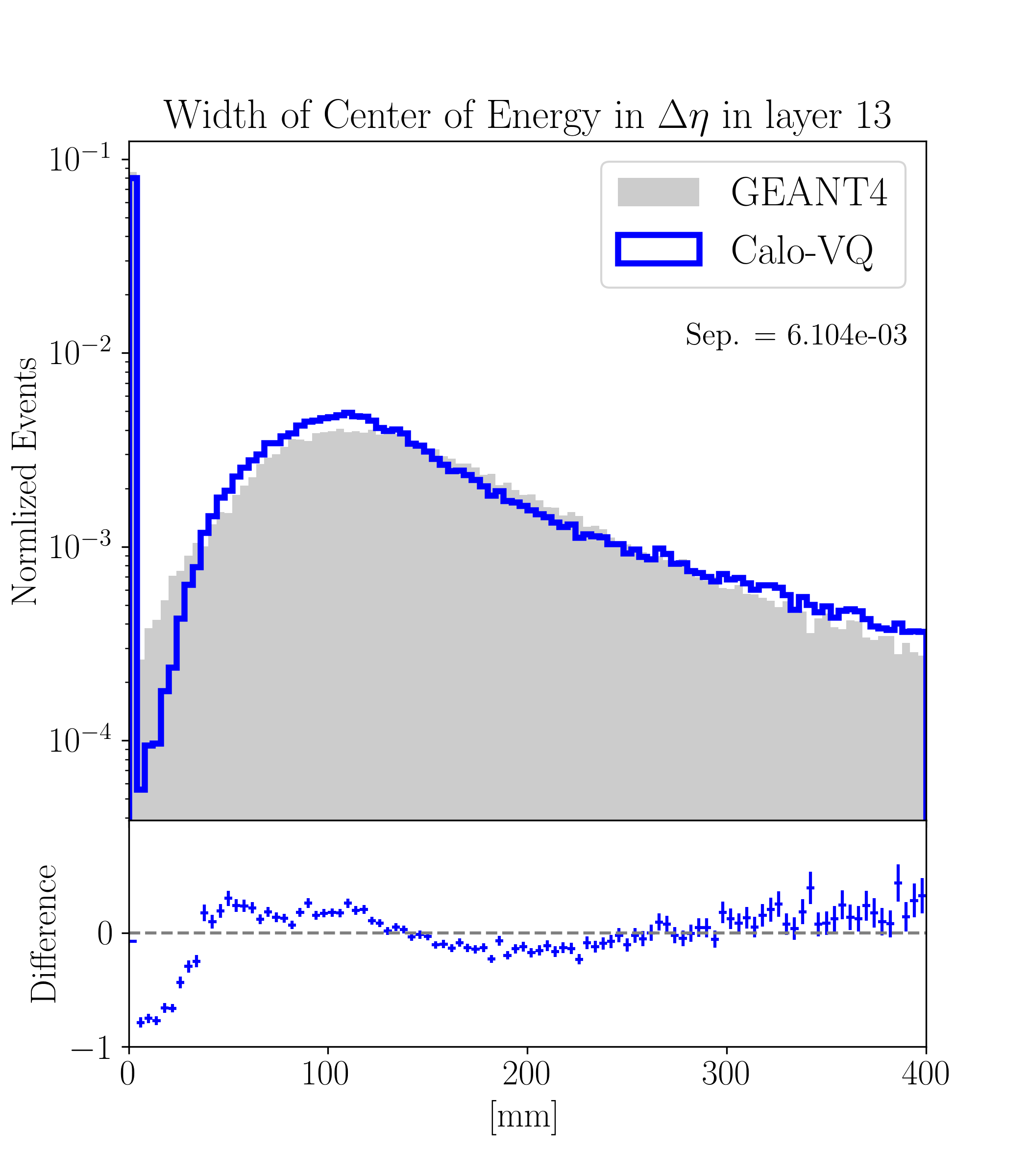}
    \includegraphics[width=.3\linewidth]{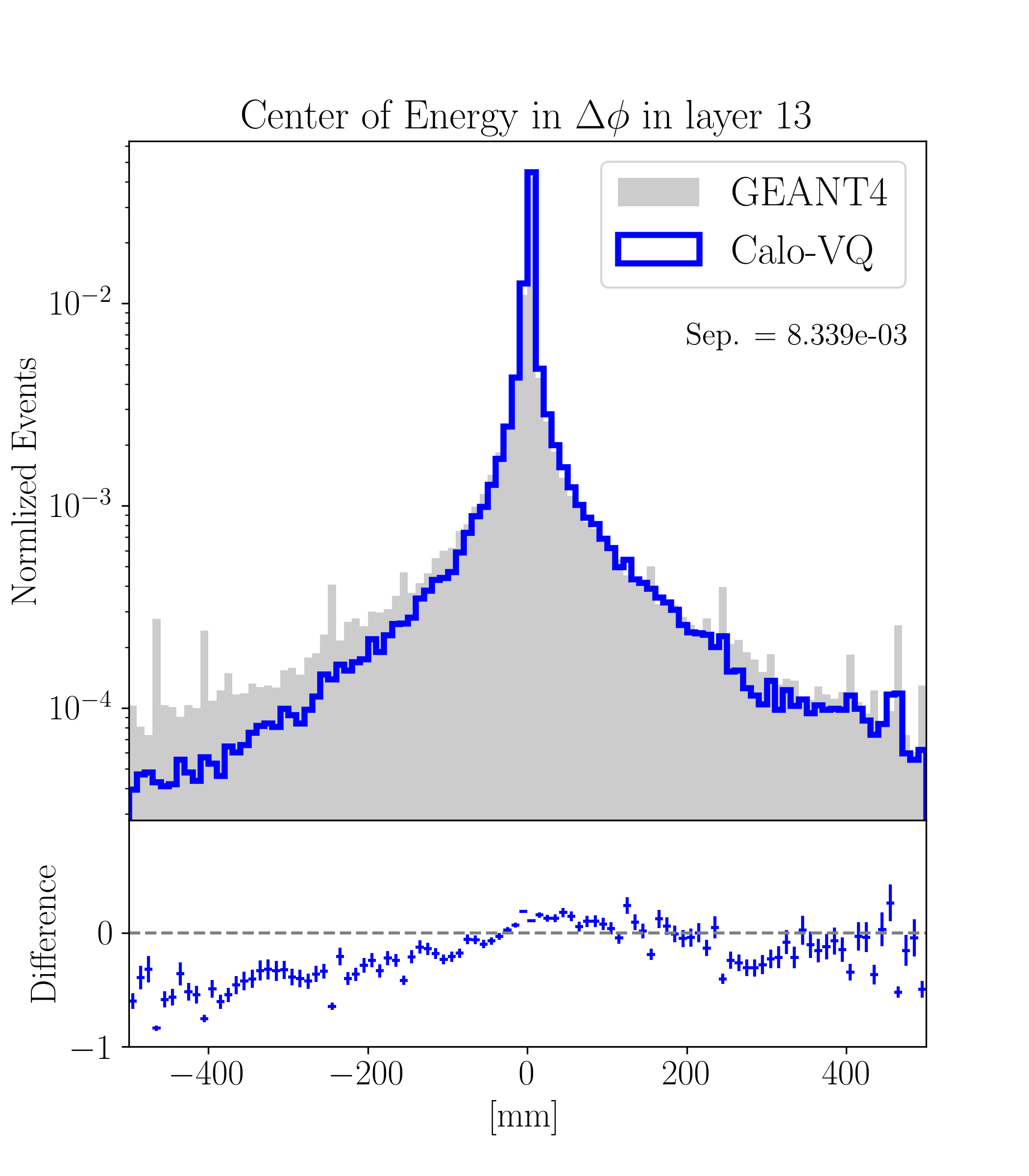}
    \includegraphics[width=.3\linewidth]{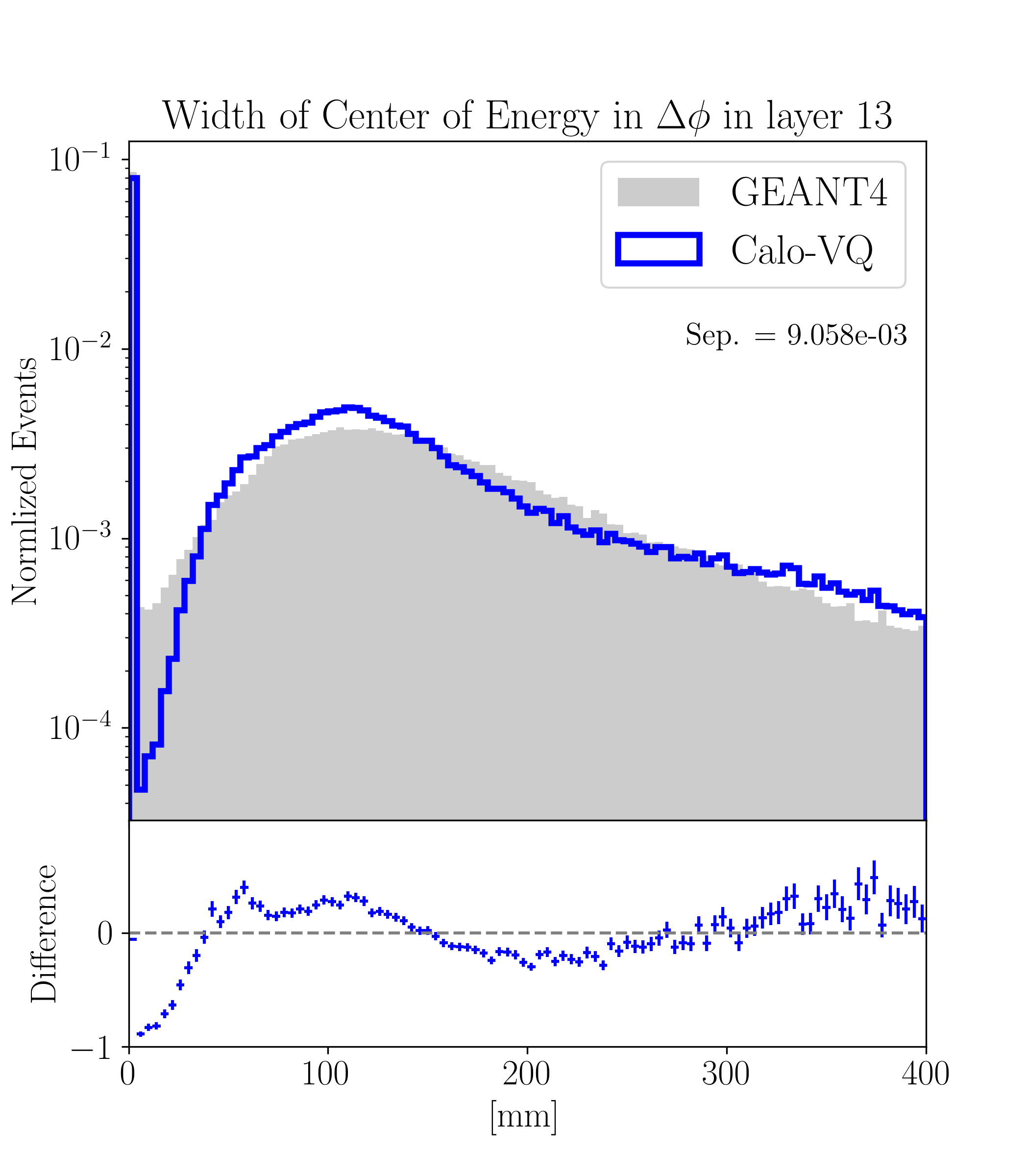}
    \caption{Distribution of physics variables from calorimeter shower for pion incident. Layer 13 is selected for demonstration. Generated is shown in orange line and reference from GEANT4 shown in solid blue. From the top left to bottom right, the physics variables are layer energy, shower center and width in $\eta$ direction, shower center and width in $\phi$ direction. }
    \label{fig:ds1_pion_v}
\end{figure}

Two types of incident particles are studied in Dataset 1: photons and pions. Models are trained and evaluated on each dataset separately. Figure~\ref{fig:ds1_pion_avg} illustrates the averaged energy distribution in each calorimeter cell across all sampled events for pion particles. Figure~\ref{fig:ds1_E} and \ref{fig:ds1_pion_v} display the distribution of individual cell energies, total energy responses, and shower variables for selected layers. The metrics of each layer are summarized in in Table~\ref{tab:photon_v}, \ref{tab:pion_v}. Generally, good agreement is observed across different incident energy conditions, with metrics for energy response better than 0.01.

\begin{table}[]\centering
\begin{tabular}{cccccc}
\hline
Sep. for Layer $K$ & $E_{K}$  & $\overline{\eta (K)}$ & $\sigma_{\eta (K)}$ & $\overline{\phi (K)}$ & $\sigma_{\phi (K)}$ \\
\hline
  0   & 0.004 & - & - & - & -  \\
  1   & 0.001 & 0.018 & 0.017 & 0.015 & 0.023 \\
  2   & 0.001 & 0.007 & 0.012 & 0.005 & 0.012  \\
  3   & 0.001 & - & - & - & - \\
 12   & 0.002 & - & - & - & - \\
\hline
\end{tabular}
\caption{Separation metrics for different layer of photon dataset. For layer 0, 3 and 12 there is no segmentation along $\eta$ and $\phi$ direction so no definition of some shower variables. }
\label{tab:photon_v}
\end{table}

\begin{table}[]\centering
\begin{tabular}{cccccc}
\hline
Sep. for Layer $K$ & $E_{K}$  & $\overline{\eta (K)}$ & $\sigma_{\eta (K)}$ & $\overline{\phi (K)}$ & $\sigma_{\phi (K)}$ \\
\hline
  0   &   0.007  & -  & -  & -  & -  \\
  1   &   0.006 & 0.015 & 0.037 & 0.027 & 0.019   \\
  2   &   0.003 & 0.015 & 0.018 & 0.015 & 0.016  \\
  3   &   0.005  & -  &  - &  - & - \\
 12   &   0.002 & 0.012 & 0.011 & 0.014 & 0.020  \\
 13   &   0.001 & 0.013 & 0.006 & 0.009 & 0.008  \\
 14   &   0.001   & -  & -  & -  & - \\
\hline
\end{tabular}
\caption{Separation metrics for different layer of pion dataset. For layer 0, 3 and 14 there is no segmentation along $\eta$ and $\phi$ direction so no definition of some shower variables. }
\label{tab:pion_v}
\end{table}

\subsection{Dataset 2: electron}
\label{ssec:ds2}
\begin{figure}[]
    \centering
    \includegraphics[width=.6\linewidth]{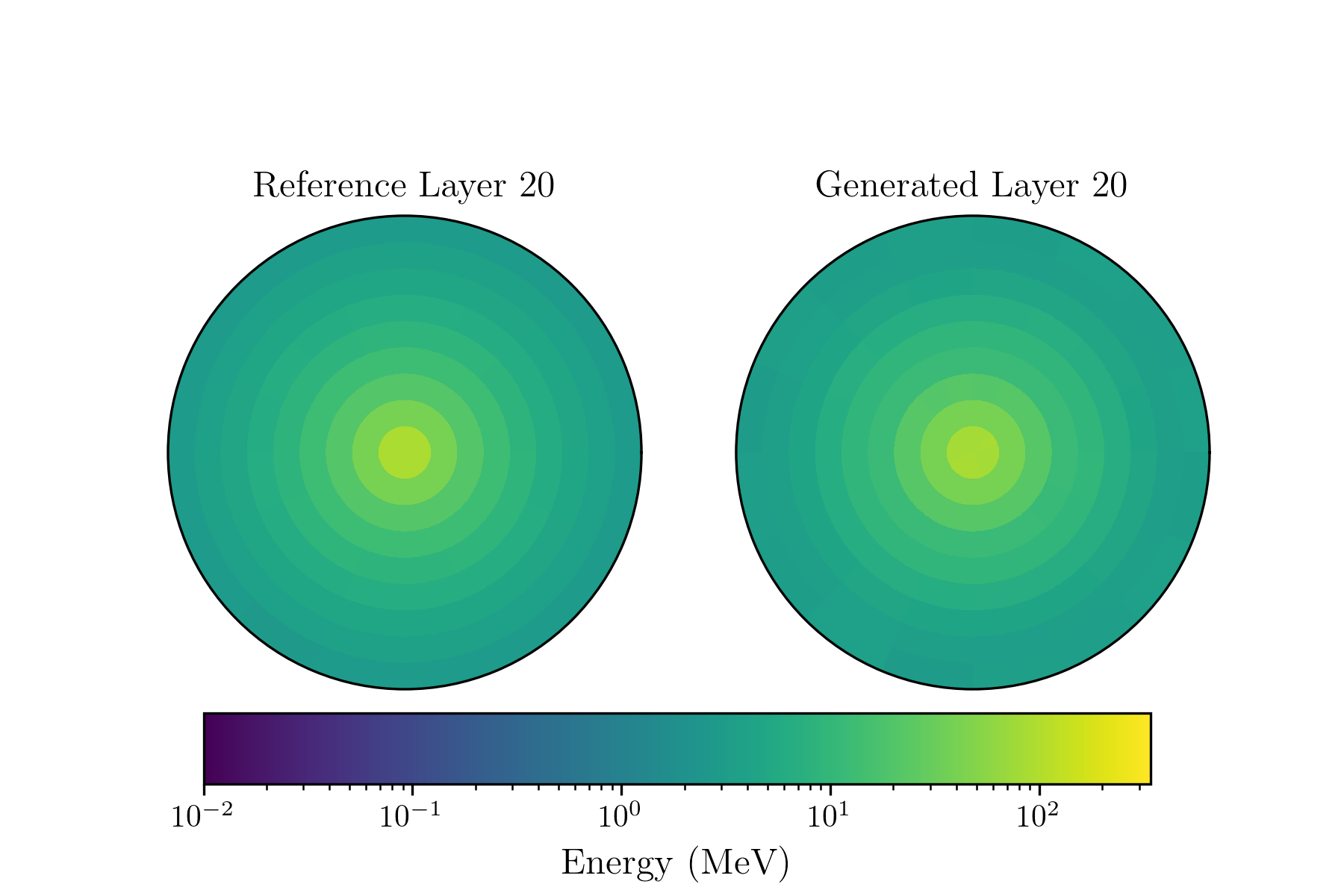}
    \caption{Average energy deposition of calorimeter cells in layer 20 of dataset2. Each shower is induced by single incident of electron. Generated on the right and reference on the left. }
    \label{fig:ds2_avg}
\end{figure}

The averaged image of all electron showers is depicted in Figure~\ref{fig:ds2_avg} for a selected calorimeter layer.

\begin{figure}[]
    \centering
    \includegraphics[width=.3\linewidth]{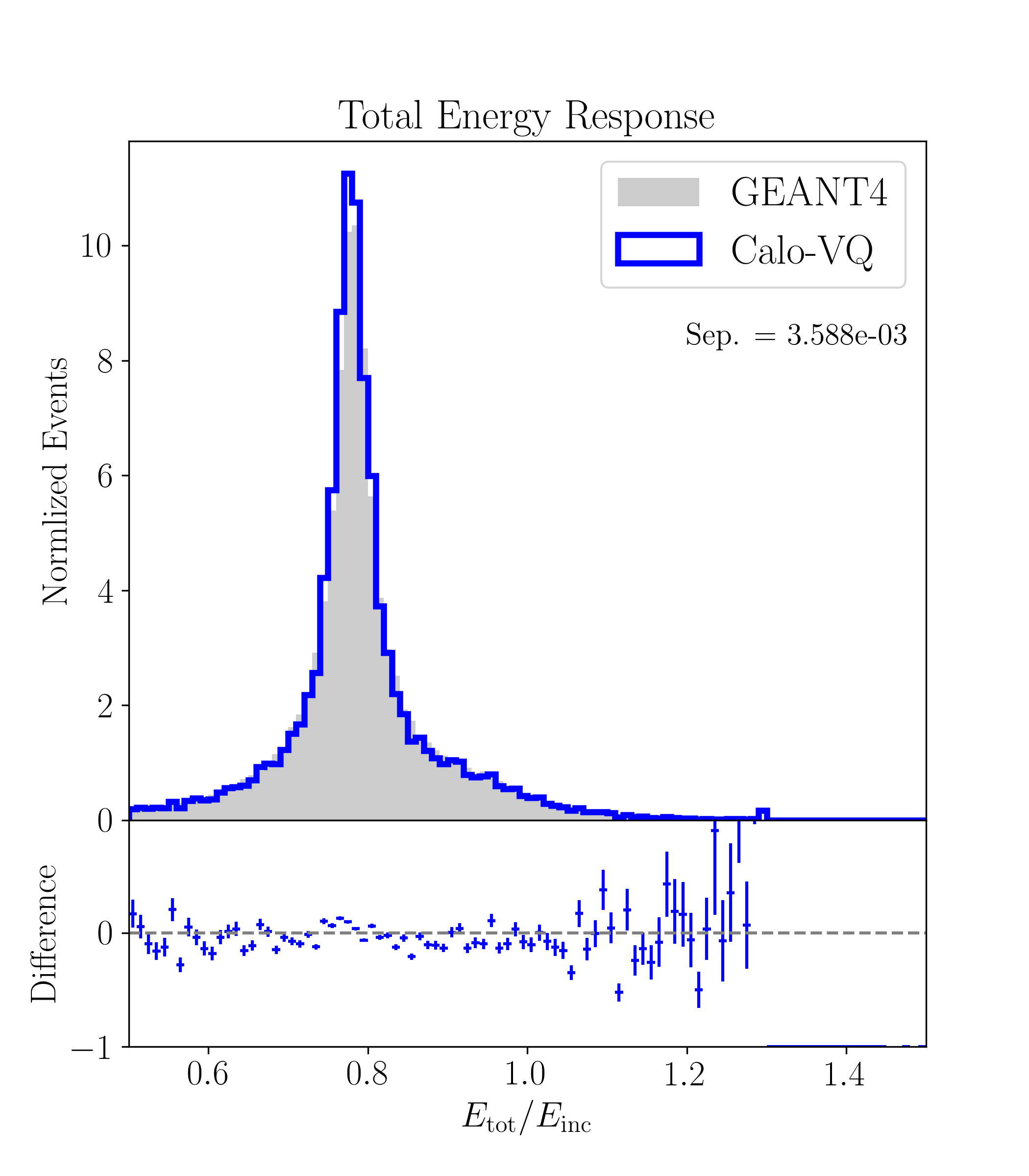}
    \includegraphics[width=.3\linewidth]{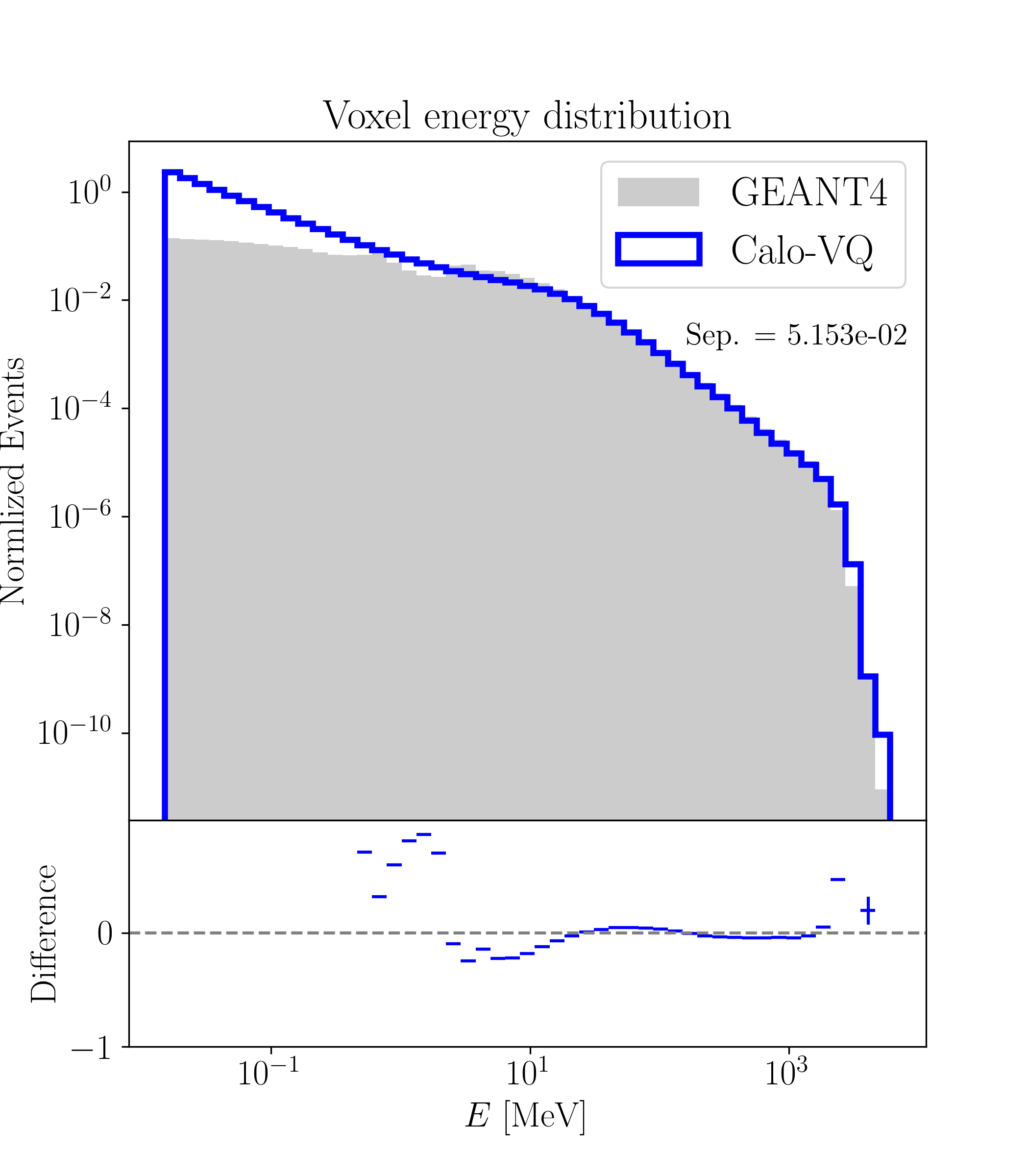} 
    \caption{Distribution of cell energy and total energy response of electron shower in dataset2. Generated is shown in orange line and reference from GEANT4 shown in solid blue.}
    \label{fig:ds2_E}
\end{figure}

\begin{figure}[]
    \centering
    \includegraphics[width=.3\linewidth]{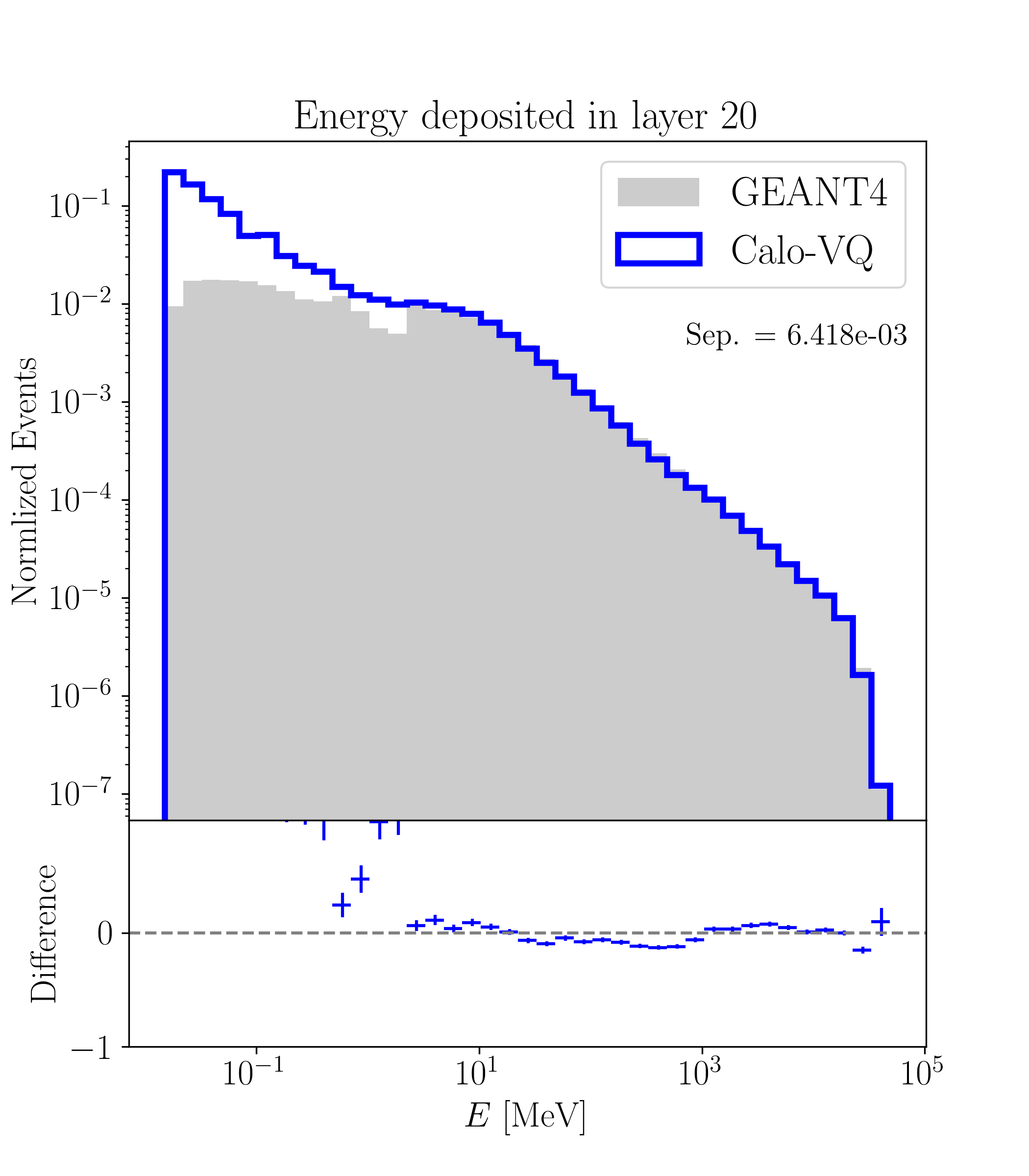} 
    \includegraphics[width=.3\linewidth]{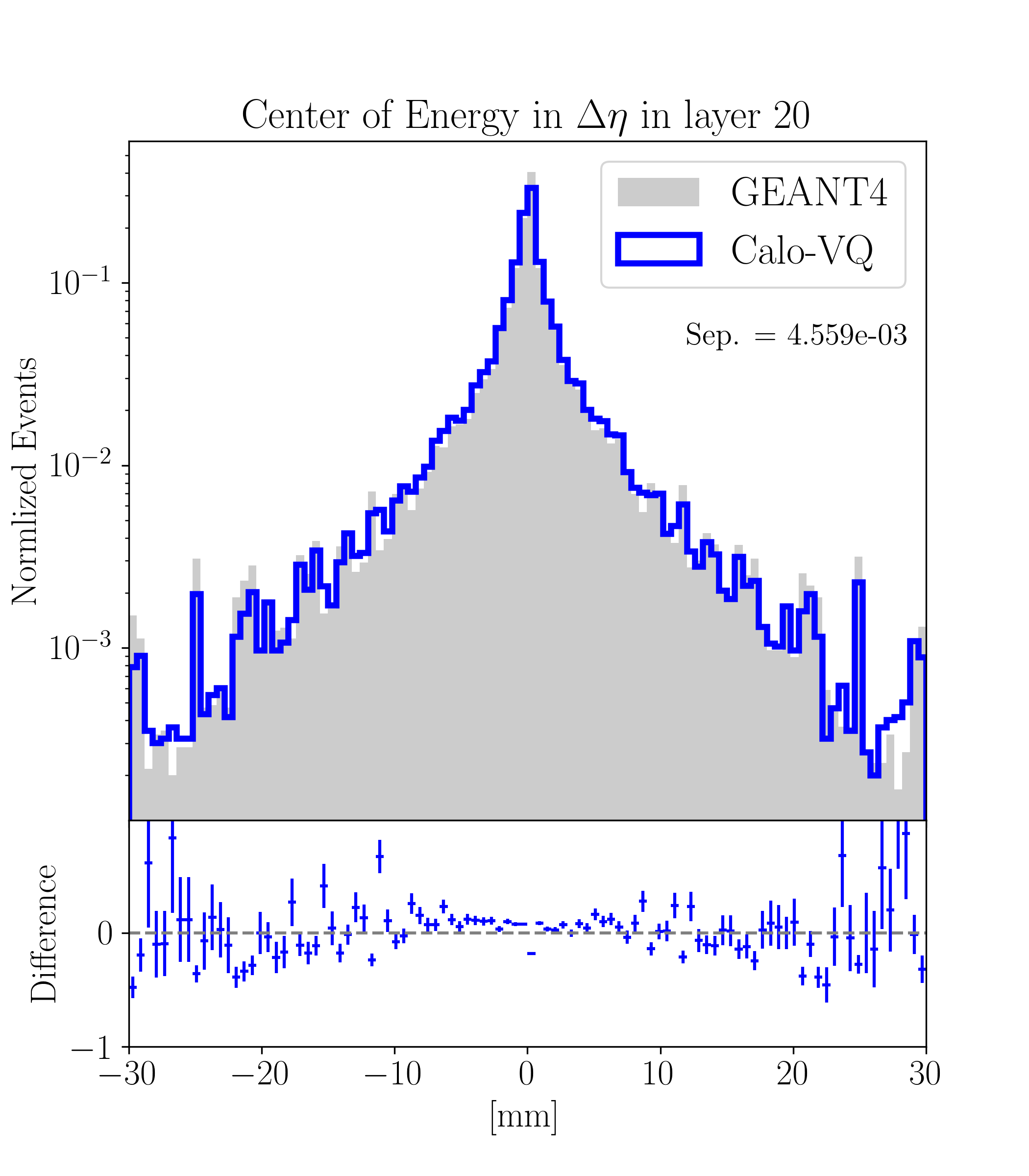}
    \includegraphics[width=.3\linewidth]{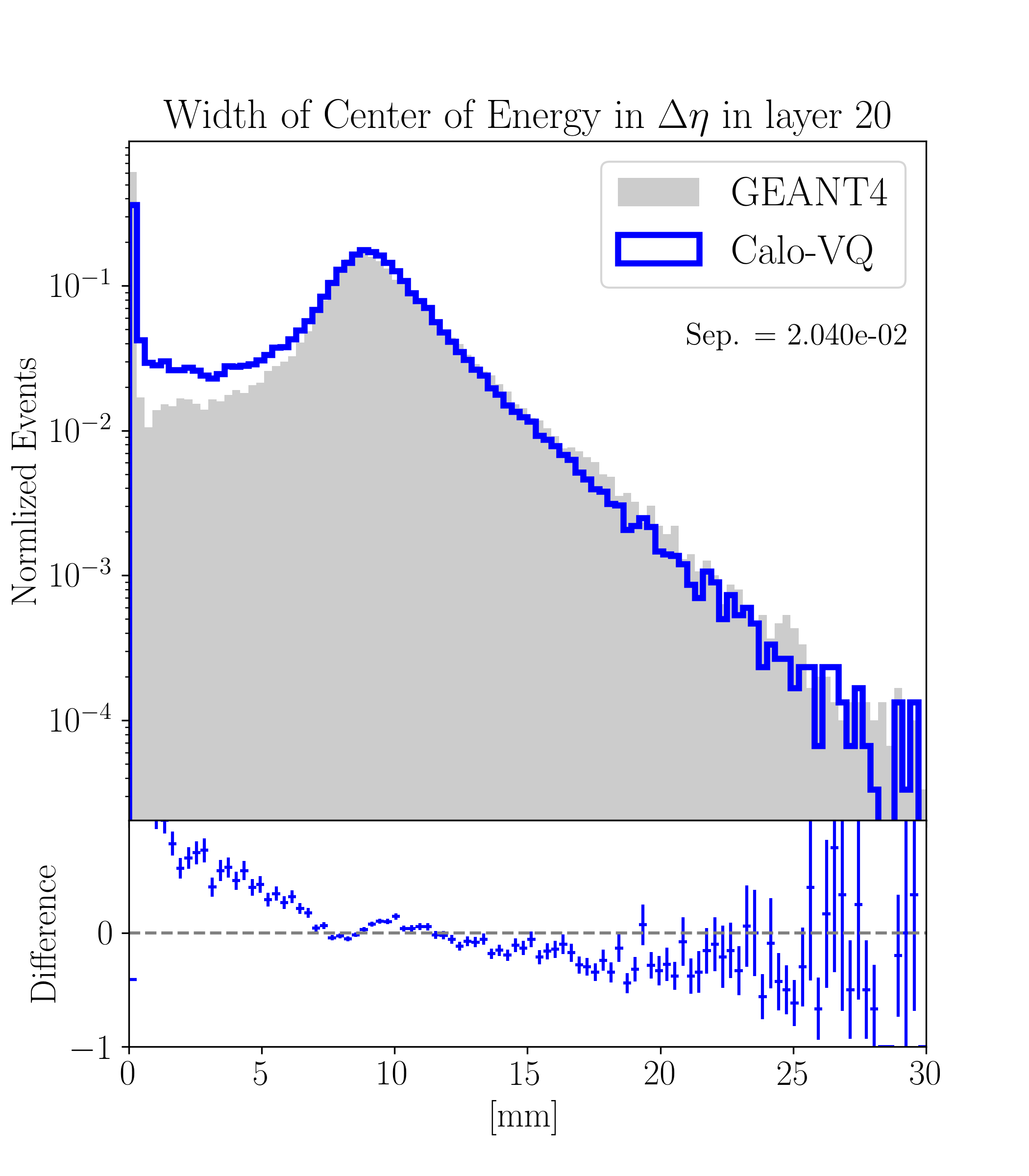}
    \includegraphics[width=.3\linewidth]{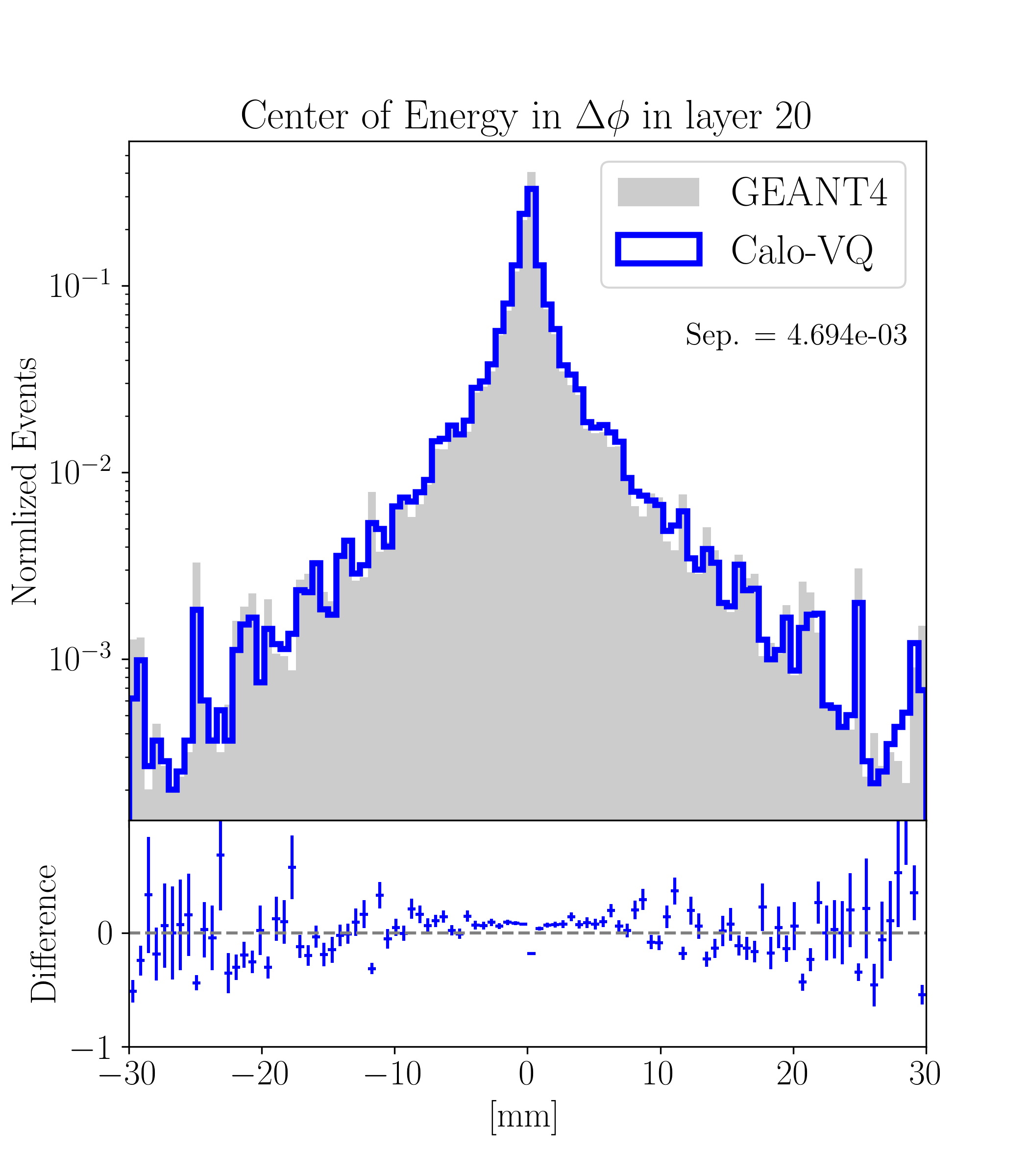}
    \includegraphics[width=.3\linewidth]{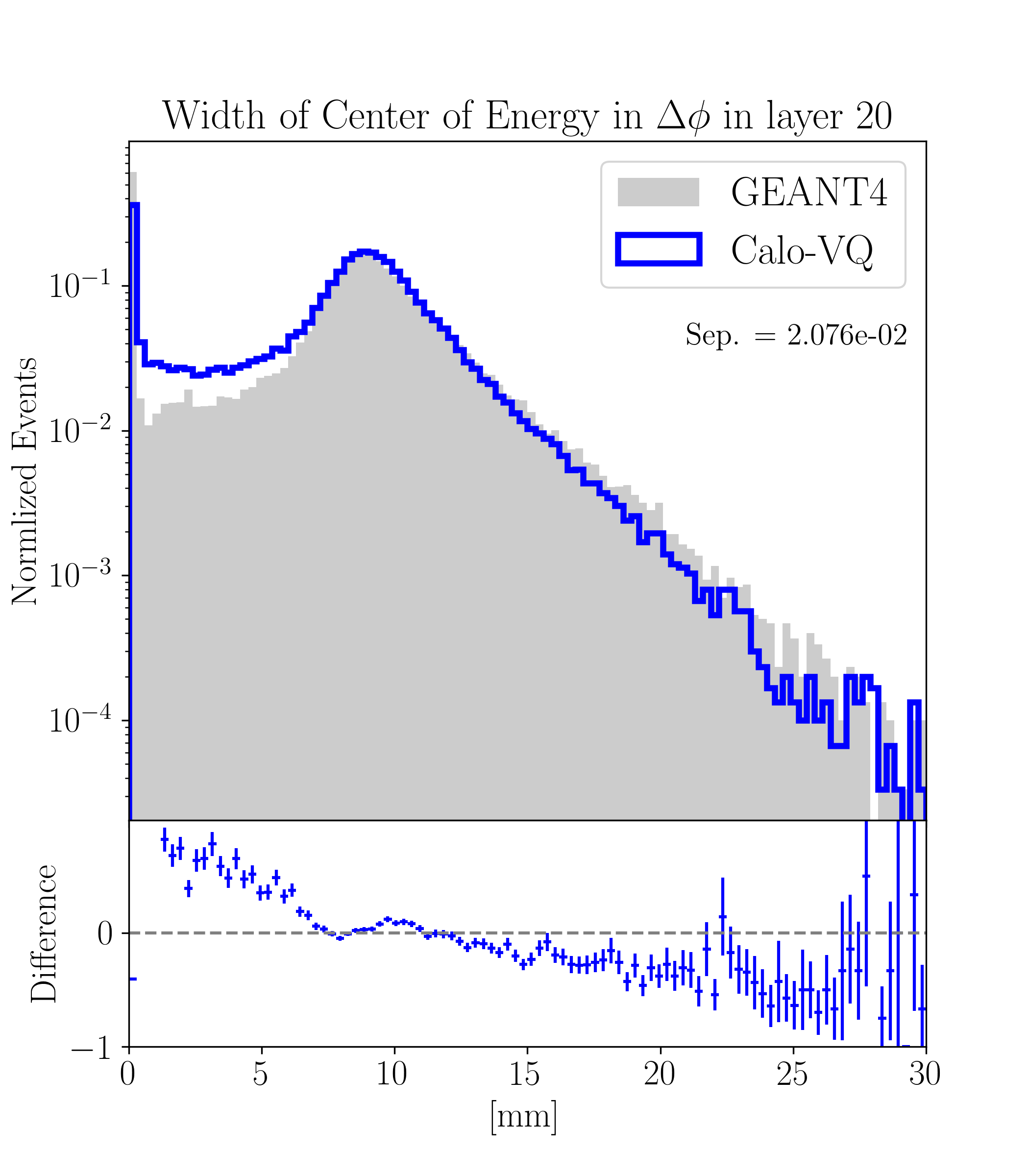}
    \caption{Distribution of physics variables from calorimeter shower for electron incident in dataset2. Layer 20 is selected for demonstration. Generated is shown in orange line and reference from GEANT4 shown in solid blue. From the top left to bottom right, the physics variables are layer energy, shower center and width in $\eta$ direction, shower center and width in $\phi$ direction.}
    \label{fig:ds2_v}
\end{figure}

\begin{figure}[]
    \centering
    \includegraphics[width=.8\linewidth]{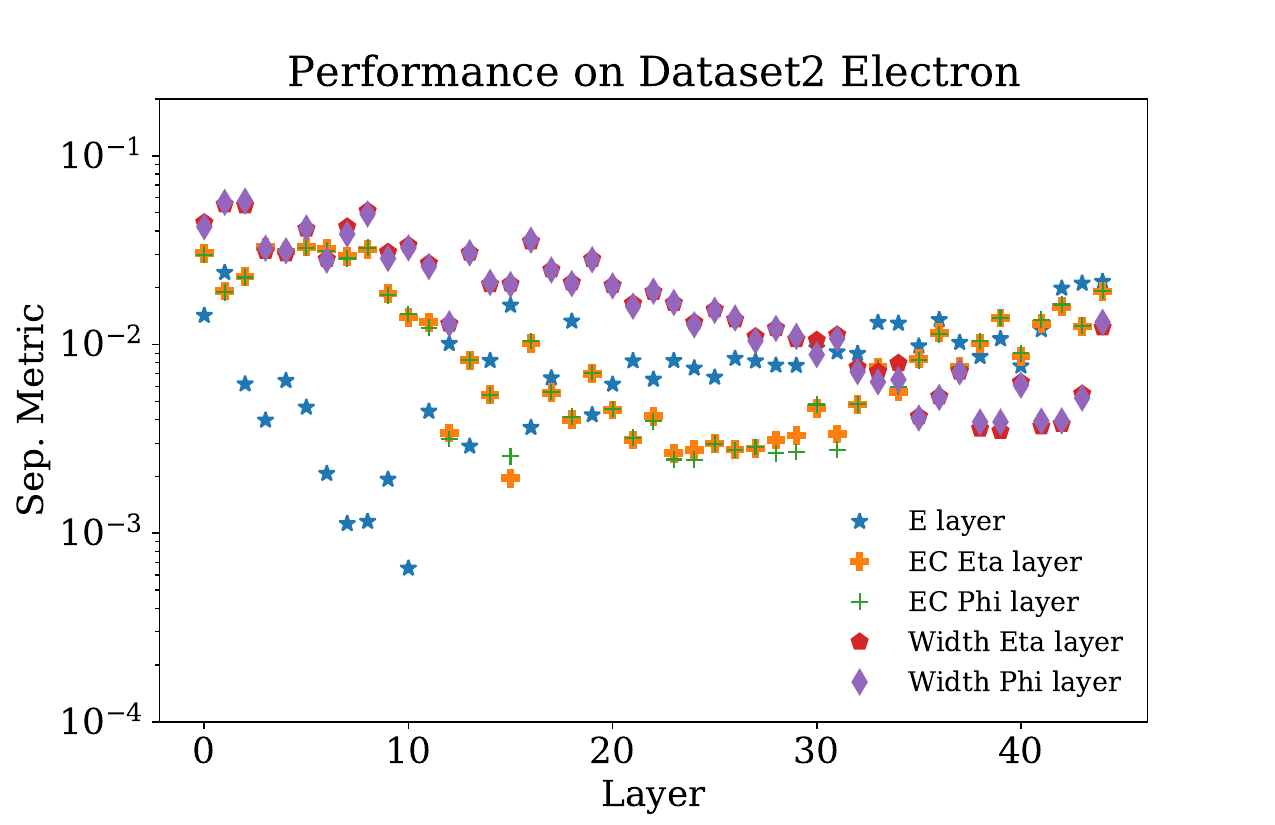}
    \caption{Separation metric of different physics variables of each layer. The lower separation means the smaller difference between fast simulation and GEANT4 simulation of calorimeter response. The definitions of variable follow those listed in Table~\ref{tab:v}.}
    \label{fig:ds2_vg}
\end{figure}

Distributions and metrics are summarized in Figures~\ref{fig:ds2_E}, \ref{fig:ds2_v} and ~\ref{fig:ds2_vg}. Despite the significantly increased granularity compared to Dataset 1, the model demonstrates strong performance in terms of modeling physics variables, with most metrics no worse than 0.01.

\subsection{Dataset 3: electron}
\label{ssec:ds3}
\begin{figure}[]
    \centering
    \includegraphics[width=.4\linewidth]{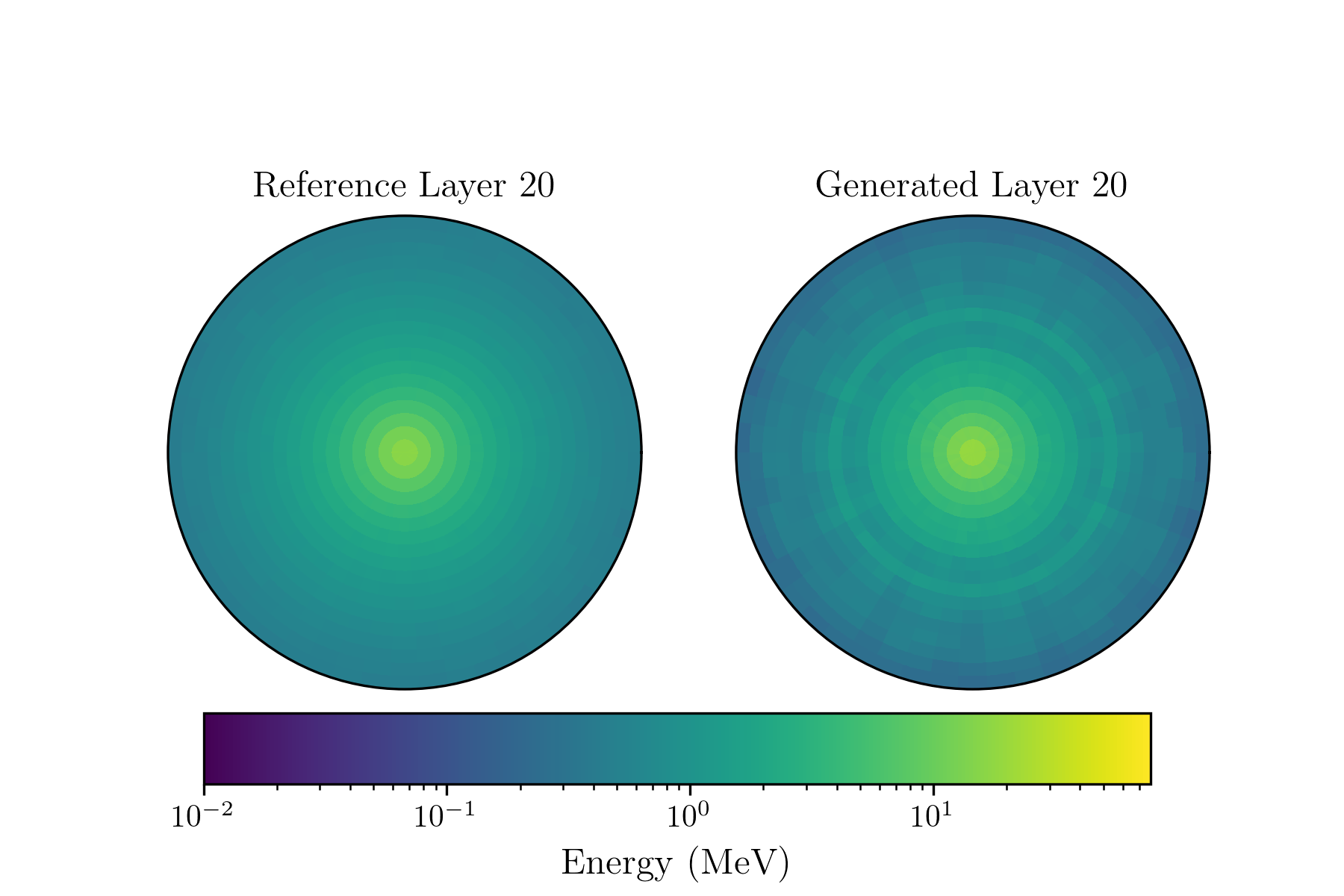} 
    \includegraphics[width=.4\linewidth]{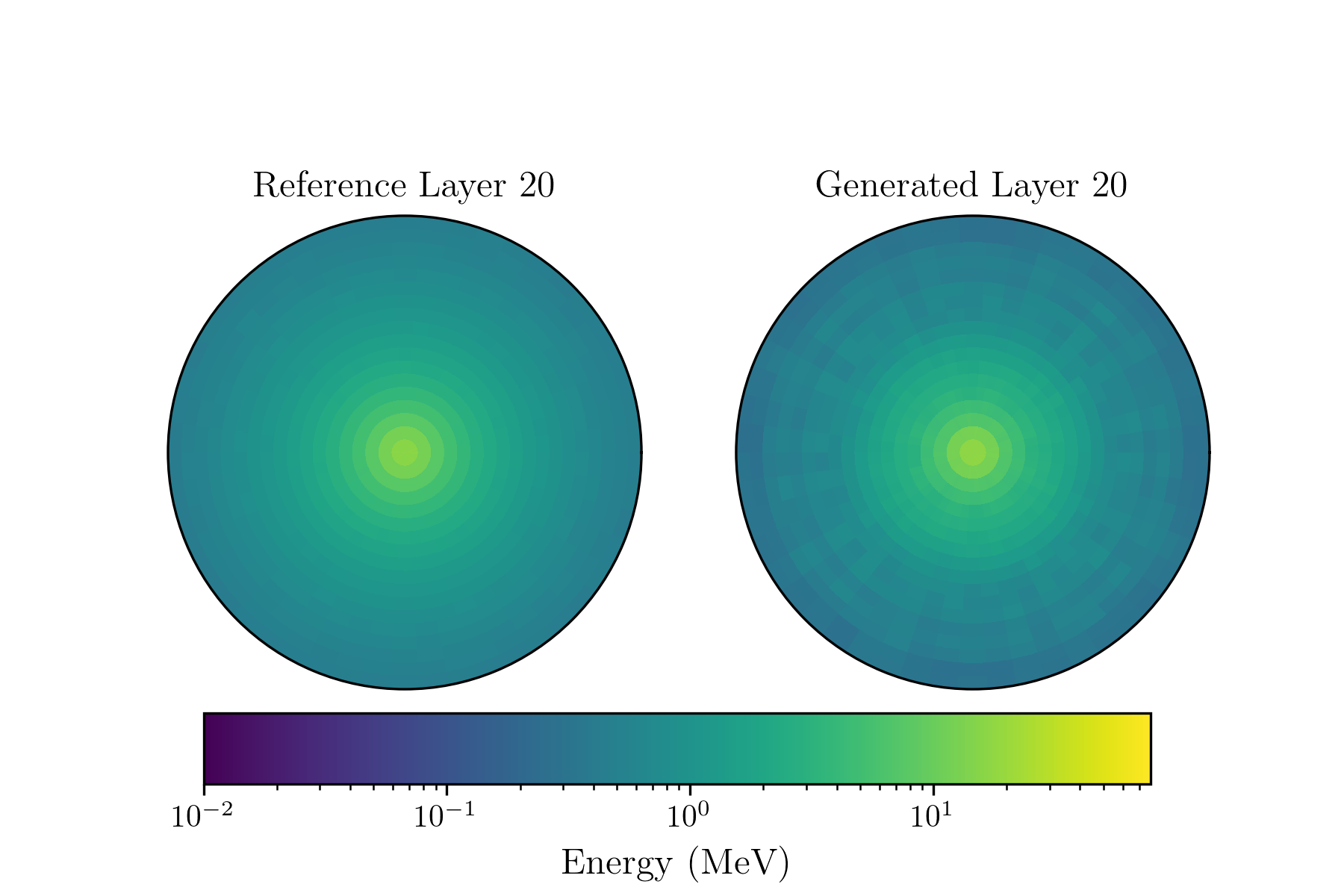}
    \caption{Average energy deposition of calorimeter cells in layer 20 of dataset3 comparing the reference and generated. Each shower is induced by single incident of electron. The result from vanilla model is shown on the left and layer-wise normalized model on the right. }
    \label{fig:ds3_avg}
\end{figure}

\begin{figure}[]
    \centering
    \includegraphics[width=.3\linewidth]{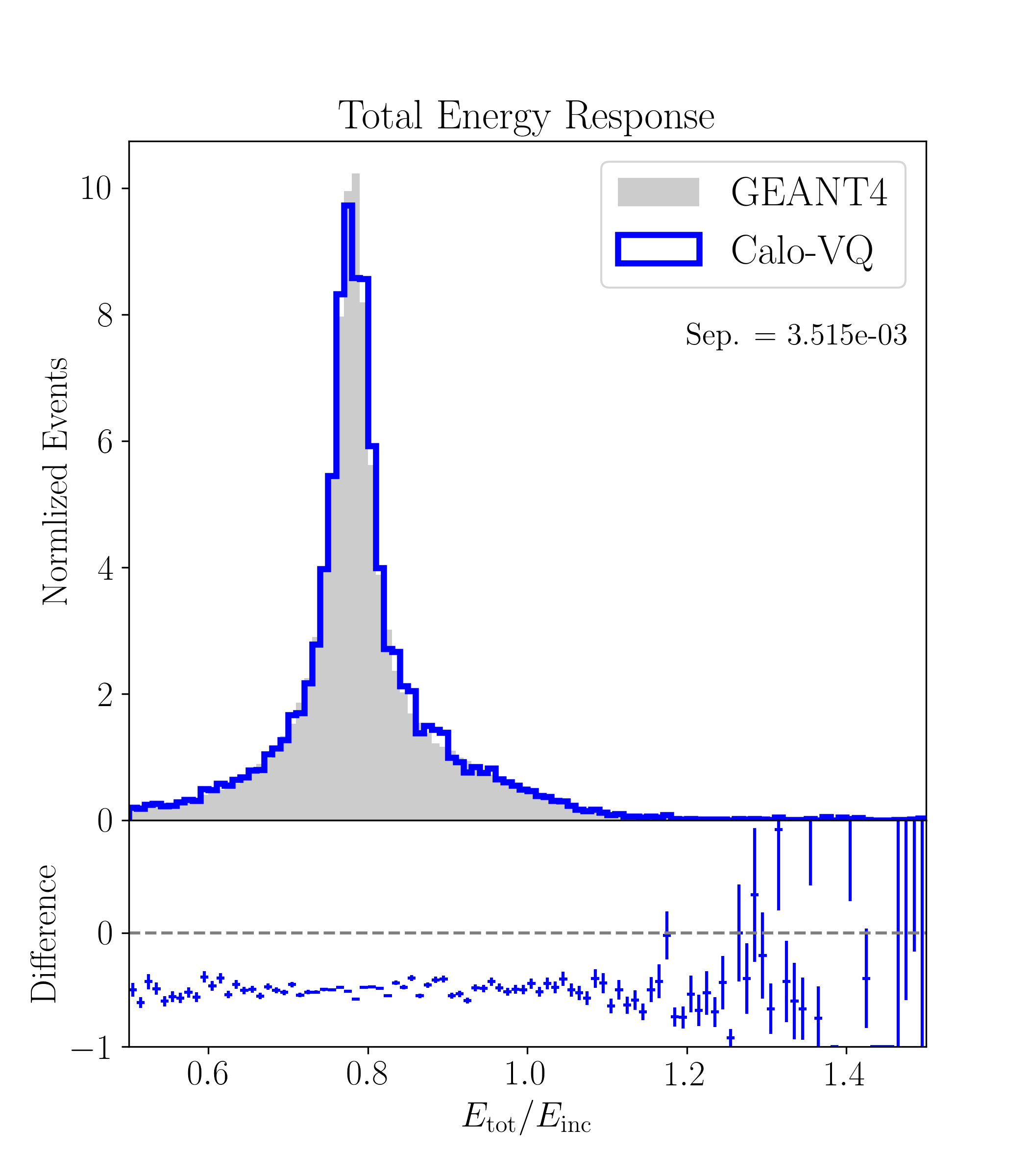}
    \includegraphics[width=.3\linewidth]{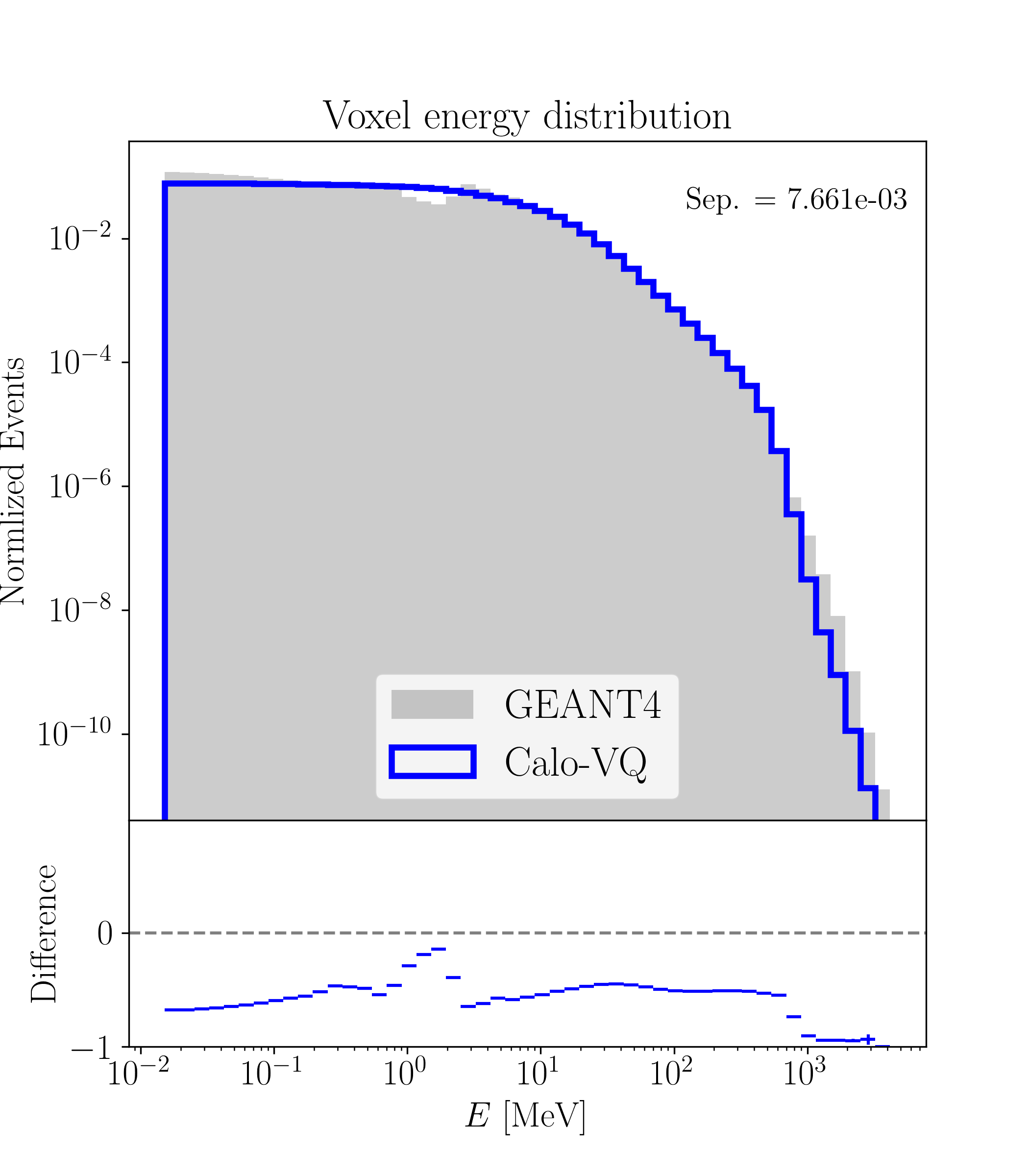} \\
    \includegraphics[width=.3\linewidth]{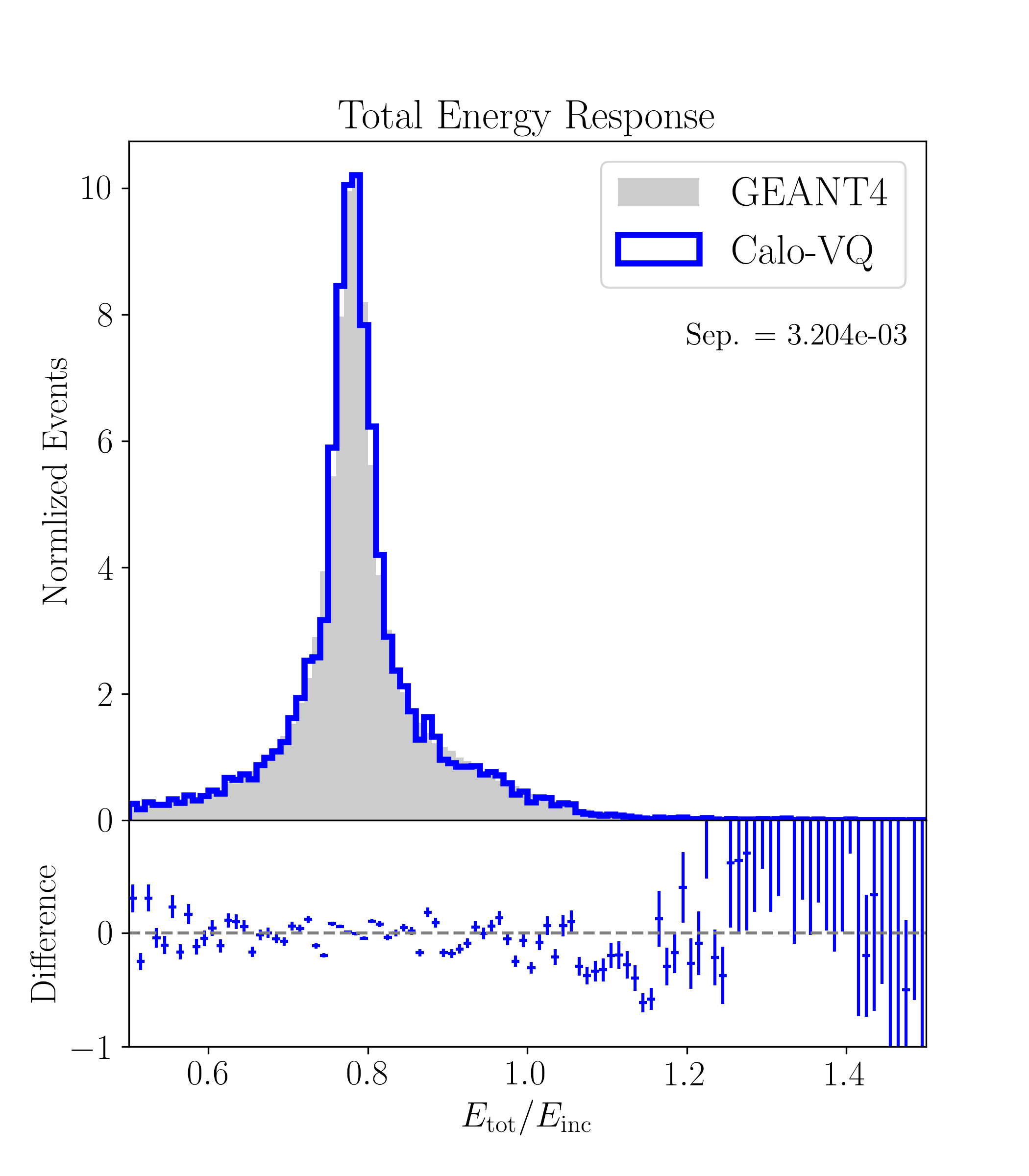}
    \includegraphics[width=.3\linewidth]{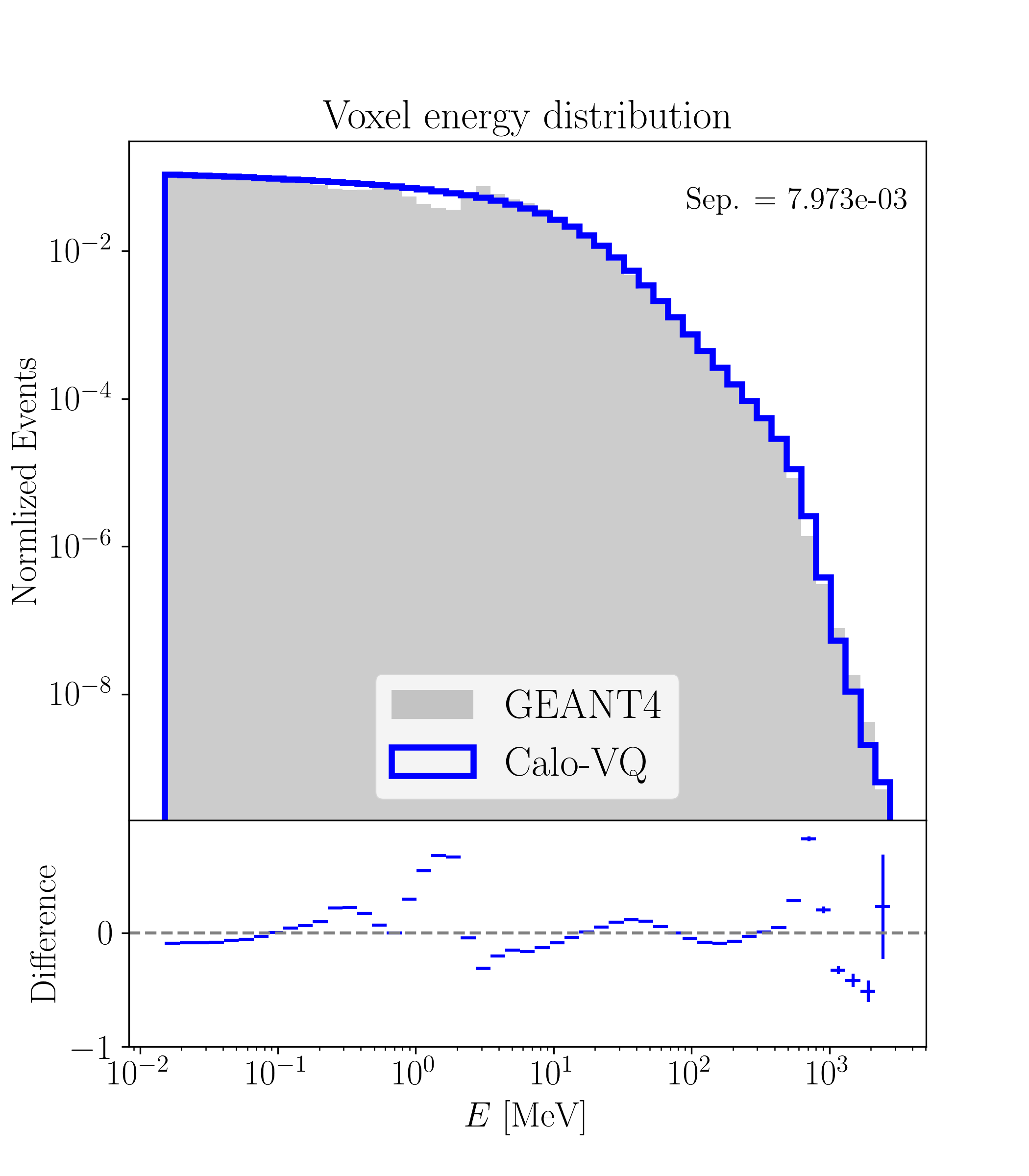} 
    \caption{Distribution of cell energy and total energy response of electron shower in dataset3. Generated is shown in orange line and reference from GEANT4 shown in solid blue. The vanilla model is shown on the top and layer-wise normalized model on the bottom.}
    \label{fig:ds3_E}
\end{figure}

\begin{figure}[]
    \centering
    \includegraphics[width=.3\linewidth]{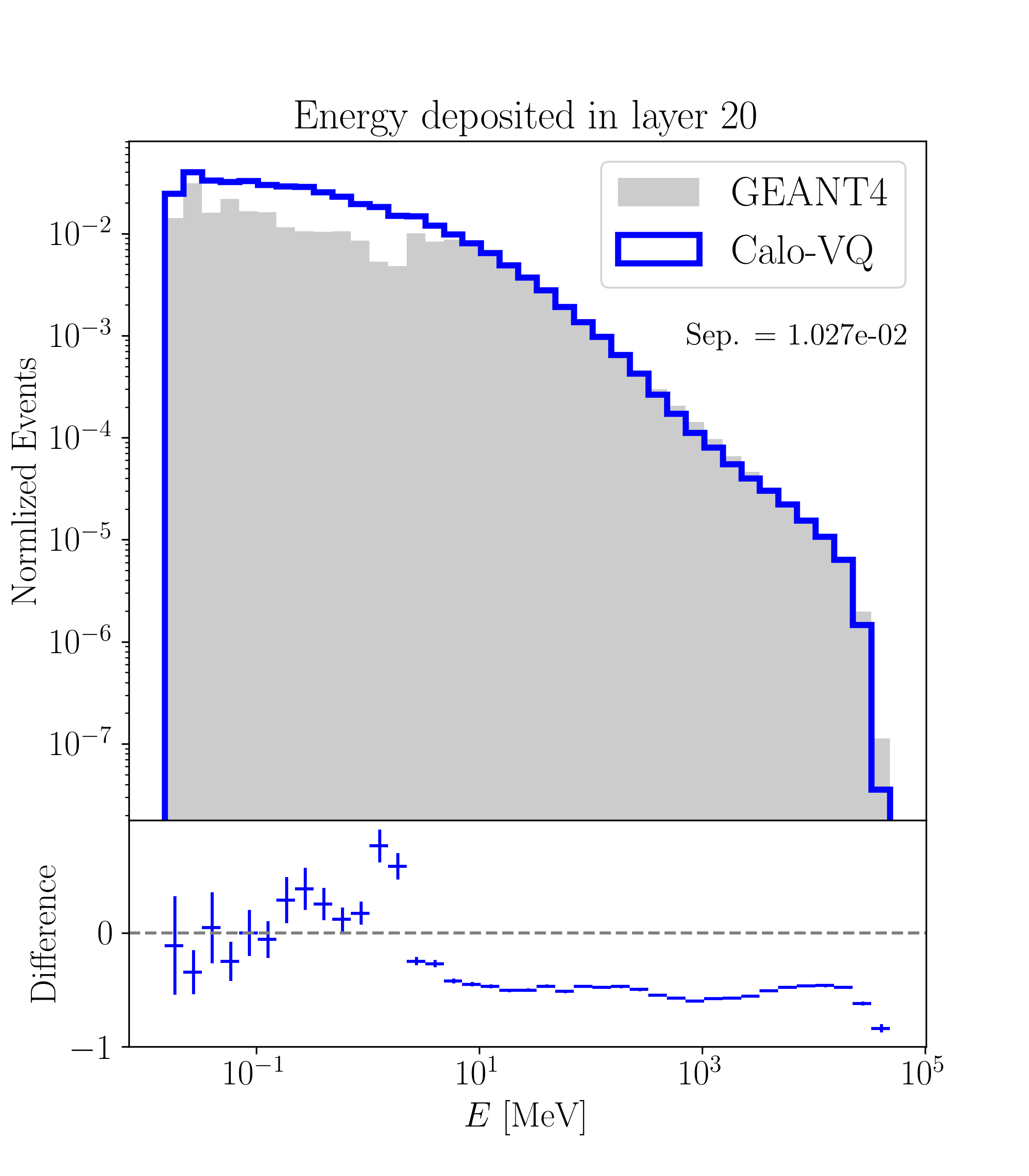} 
    \includegraphics[width=.3\linewidth]{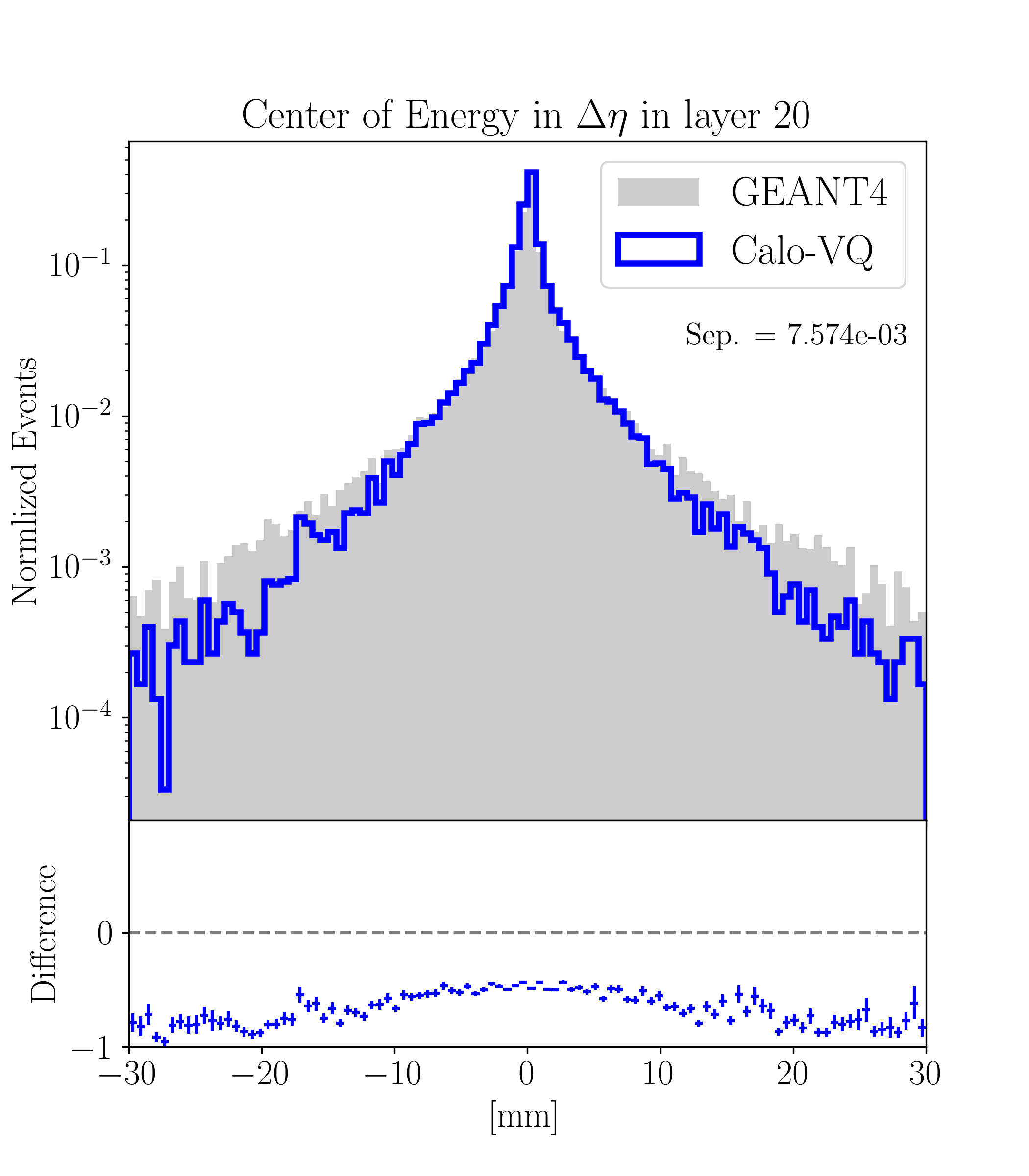}
    \includegraphics[width=.3\linewidth]{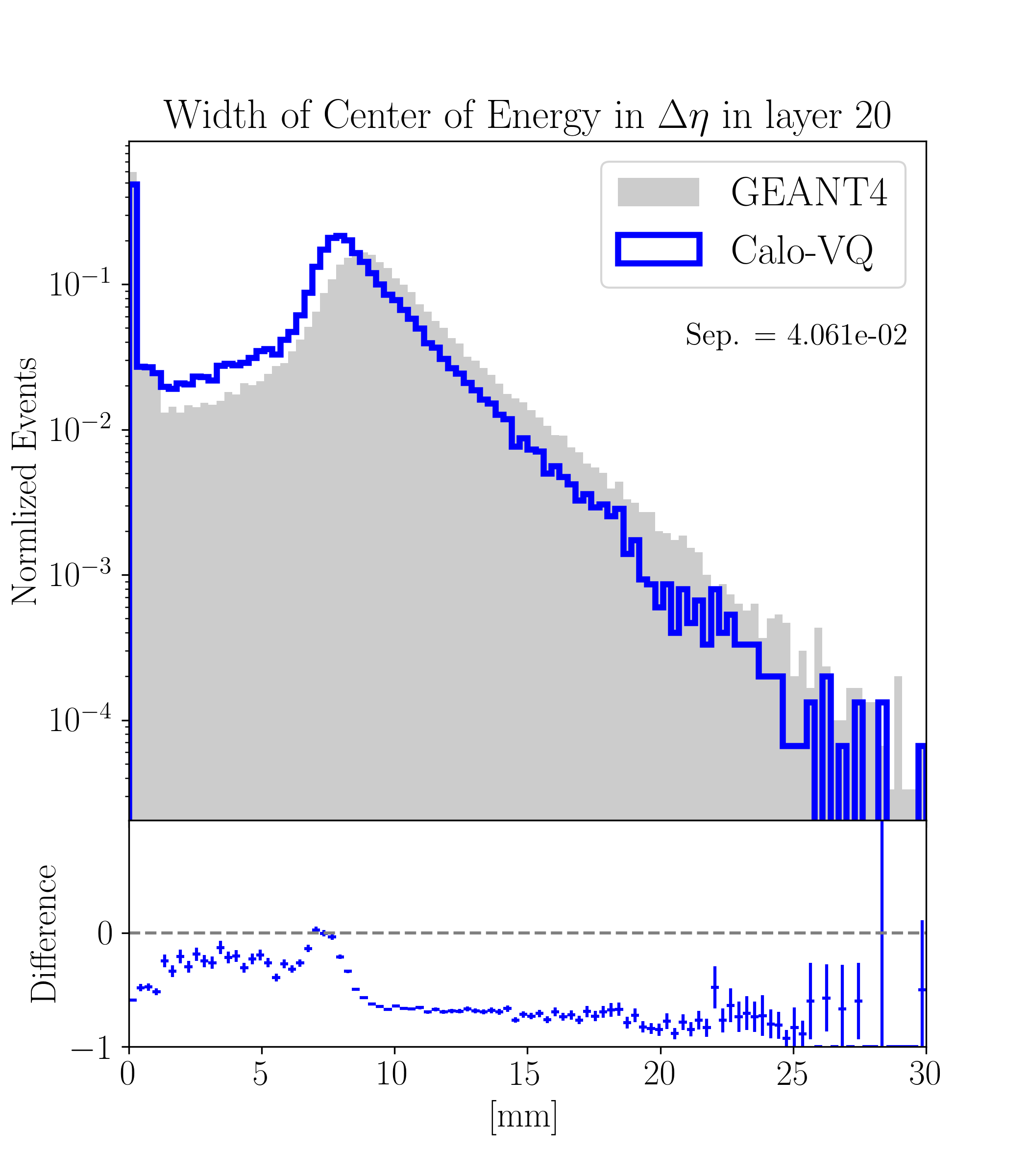} \\
    \includegraphics[width=.3\linewidth]{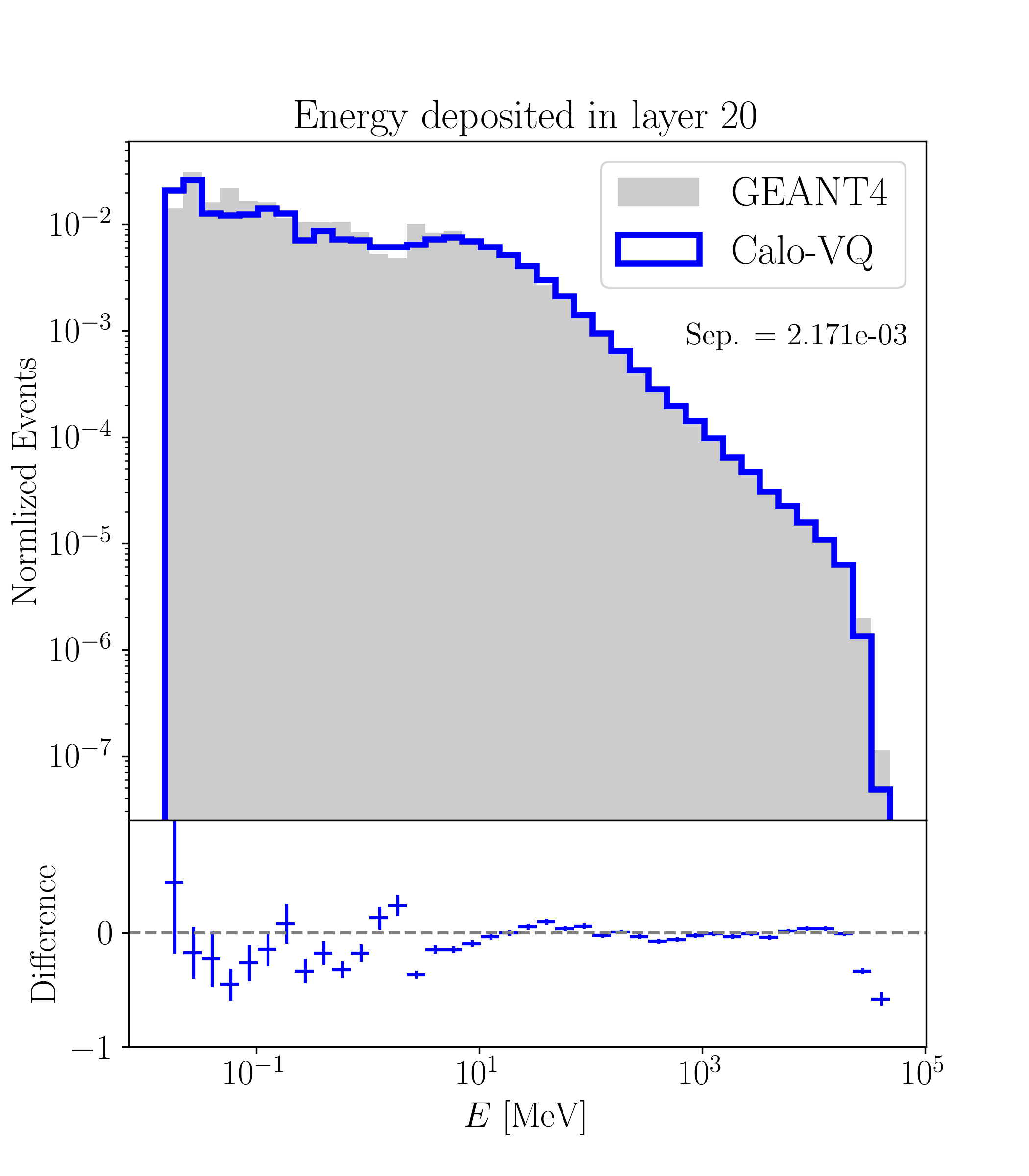} 
    \includegraphics[width=.3\linewidth]{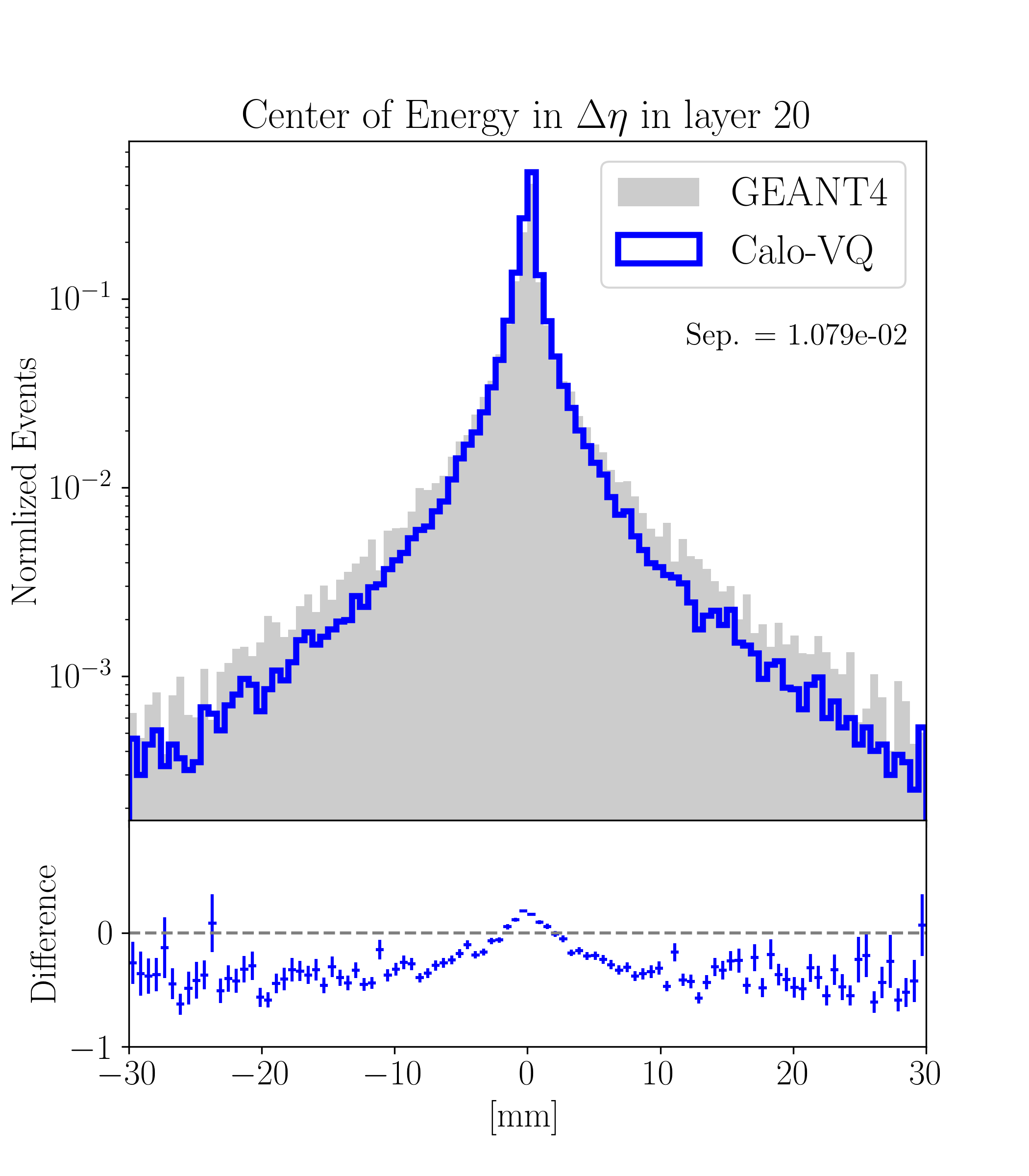}
    \includegraphics[width=.3\linewidth]{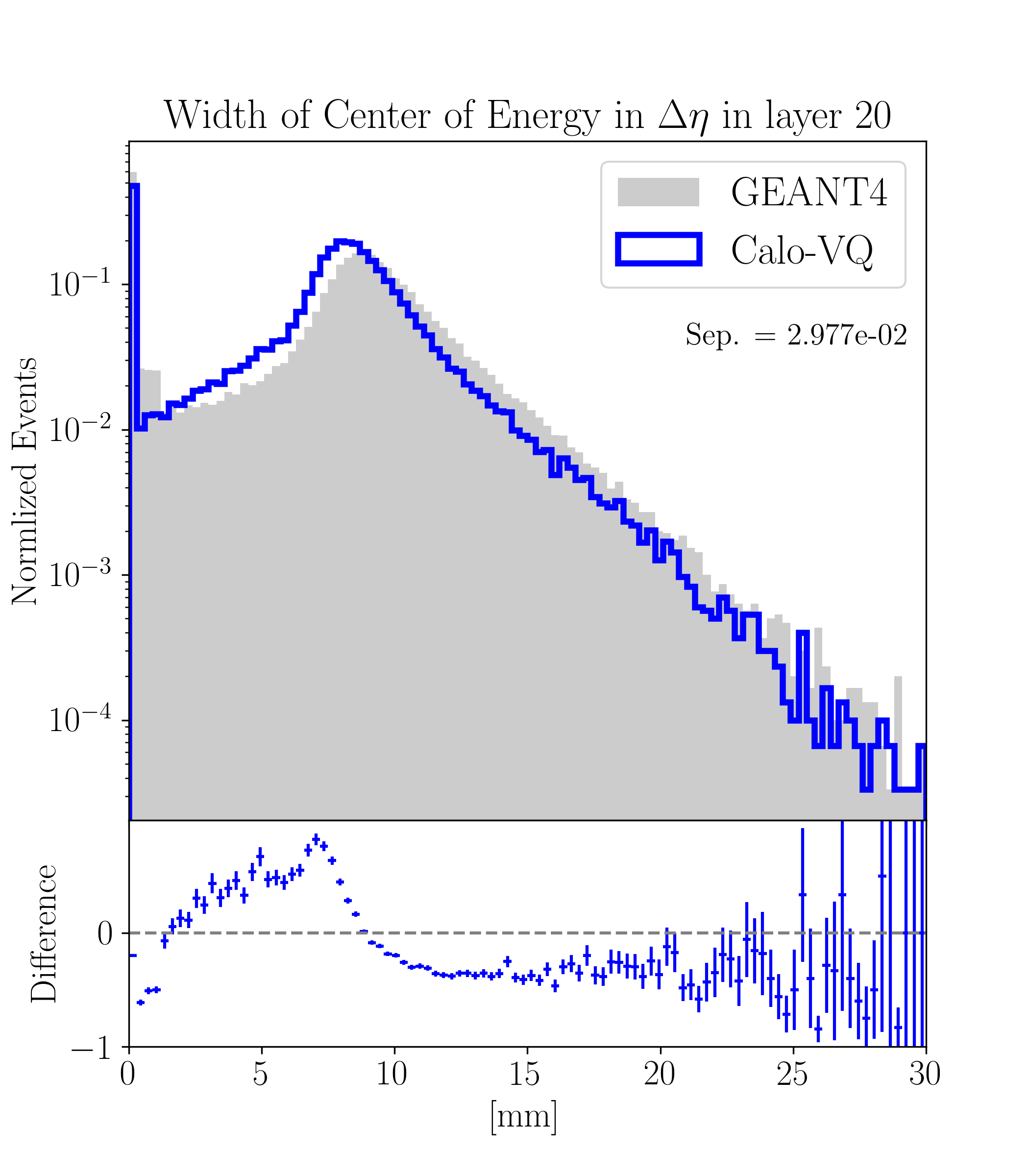}
    \caption{Distribution of physics variables from calorimeter shower for electron incident in dataset3. Layer 20 is selected for demonstration. Generated is shown in orange line and reference from GEANT4 shown in solid blue. From the top left to bottom right, the physics variables are layer energy, shower center and width in $\eta$ direction, shower center and width in $\phi$ direction. he vanilla model is shown on the top and layer-wise normalized model on the bottom.}
    \label{fig:ds3_v}
\end{figure}

\begin{figure}[]
    \centering
    \includegraphics[width=.8\linewidth]{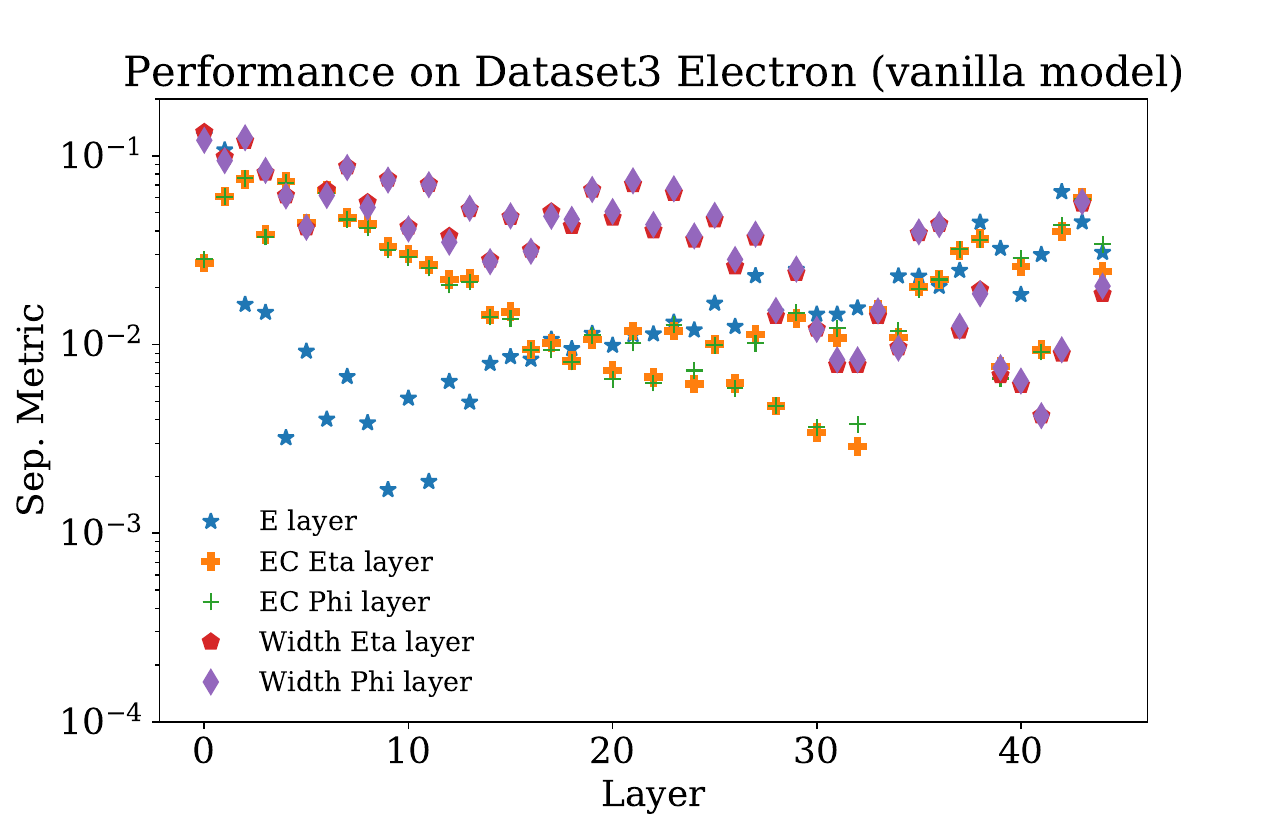}
    \caption{Separation metric of different physics variables of each layer. The lower separation means the smaller difference between fast simulation and GEANT4 simulation of calorimeter response. The definitions of variable follow those listed in Table~\ref{tab:v}.}
    \label{fig:ds3_vg}
\end{figure}

\begin{figure}[]
    \centering
    \includegraphics[width=.8\linewidth]{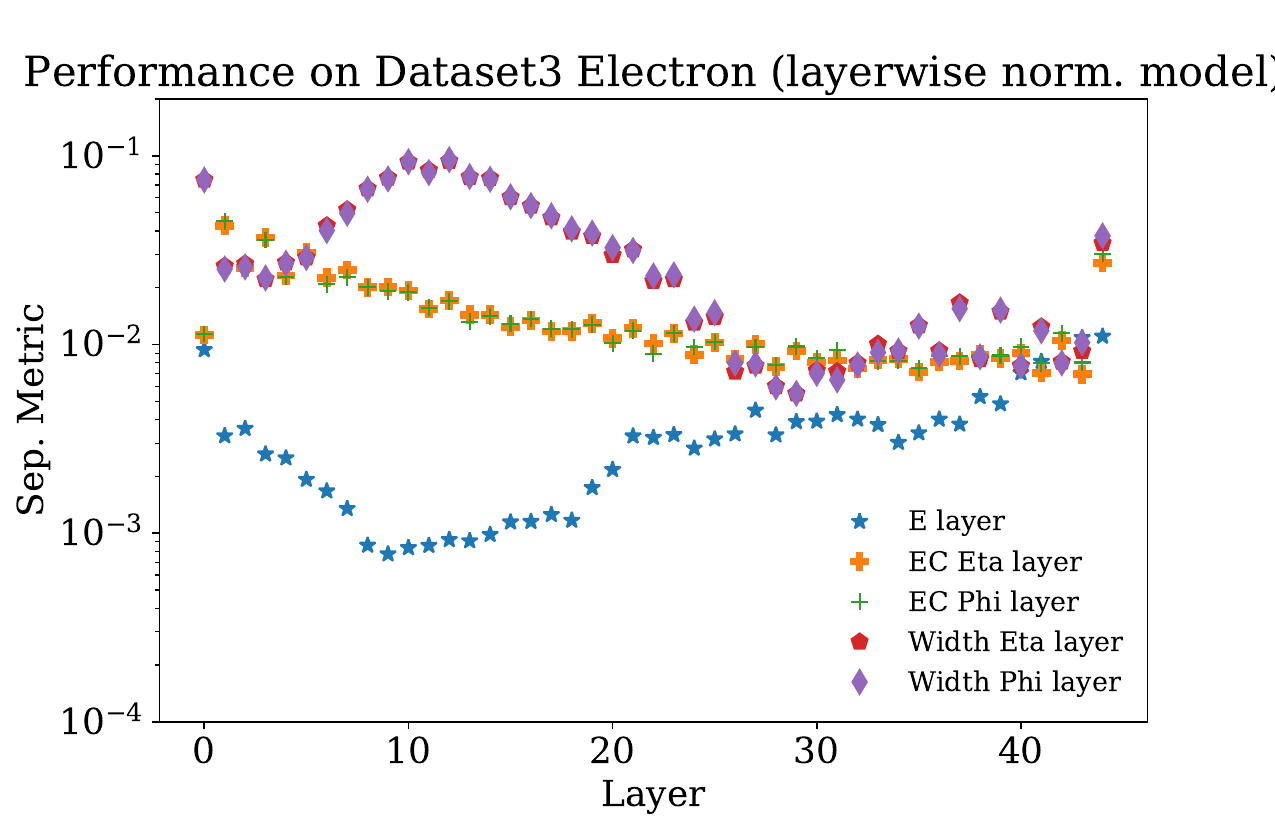}
    \caption{Separation metric of different physics variables of each layer. The lower separation means the smaller difference between fast simulation and GEANT4 simulation of calorimeter response. The definitions of variable follow those listed in Table~\ref{tab:v}.}
    \label{fig:ds3_vg2}
\end{figure}

The images, distributions, and metrics are summarized in Figures~\ref{fig:ds3_avg}, \ref{fig:ds3_E}, \ref{fig:ds3_v}, ~\ref{fig:ds3_vg} and ~\ref{fig:ds3_vg2}.  The architecture of the two-stage model can be adapted to different auto-encoder backbones optimized for different geometries. This flexibility allows for handling and maintaining performance with up to 40,500 channels, employing a similar architecture as designed for Dataset 2 just with simple scaling. This underscores the robustness and scalability of the design.

Two models are presented here, one with and one without layer-wise normalization as mentioned in Section~\ref{ssec:pre}. It is observed that layer-wise normalization improves performance, particularly in the distribution of energy sums for each layer. However, this improvement comes at the cost of increased latent size, resulting in slower generation.

\subsection{Summary Table}
\label{ssec:cost}
The number of parameters and generation time per shower for each model, along with the worst and best metric across all layers, are summarized in Table~\ref{tab:cost}. The evaluation was conducted using a single NVIDIA V100 GPU with 512 showers per batch. 
For comparison, the generation time per shower for the standard method, based on GEANT4, varies widely depending on the energy, taking approximately $O(1)-O(100)$ seconds for Dataset1~\cite{ATL-SOFT-PUB-2020-006} and around $O(100)$ seconds for Dataset 2 and 3 on average \cite{2305.11934}.
The proposed fast simulation method achieves a speedup of over 2000 times. The performance is consistently maintained across datasets with varying granularity, scaling well up to 40,500 dimensions.

Furthermore, the study reveals distinct scaling behaviors between the two stages of the model. The first stage model exhibits nearly constant scaling with different dataset complexities, demonstrating the scalability of deep convolutional networks for compression tasks. Conversely, the second stage model is only sensitive to the size of the latent space and demonstrates good generalization across different datasets. The generation quality is primarily limited by the first stage model and it impacts the computing costs solely through the latent size. Details of the performance and computing cost for both the first and second models are tabulated in Appendix~\ref{asec:cost_s12}

\begin{table}[]
\centering
\begin{tabular}{cccccc}
\hline
Dataset(Model) & Channel (D) & Latent Size (L)  & N\#Pars/M & Total time/ms & Metric(worst/best) \\
\hline
ds1(photon)   & 368       & 42               & 4.06             & 0.25 &  0.023/0.001       \\
ds1(pion)     & 533       & 46               & 4.31             & 0.28 &  0.037/0.001        \\
ds2           & 6480      & 70               & 3.32             & 0.63 &  0.057/0.001       \\
ds3(vanilla)   & 40500     & 184              & 2.16             & 1.14  &  0.133/0.002      \\
ds3(LN)    & 40500     & 274              & 2.77             & 36.10 &   0.095/0.001    \\
\hline
\end{tabular}
\caption{Number of parameters and generation time per shower of each model for different datasets. "LN" denotes the layer-wise normalized model.}
\label{tab:cost}
\end{table}

\clearpage
%%%%%%%%%%%%%%%%%%%%%%%%%%%%%%%%%%%%%%%%%%%%%%%%%%%%%%%%%%%%%%%%%%%%%%%%%%%%%%%%%%%%%%%%%%%%%%%%%%%%%%%%%%%%%%%
\section{Conclusion}
\label{sec:conclusion}
The paper introduces a novel two-stage vector-quantization based generative model for calorimeter simulation, presenting its technical framework, physics performance, and computational considerations.

Significantly, our model achieves a speedup exceeding 2000 times compared to conventional calorimeter simulation methods, while maintaining discrepancies, measured by separation metric, in total energy and energy per layer at 0.01 level and shower shape at 0.1 level. Although our model's performance does not surpass existing generative models based on normalizing flow or diffusion, it introduces a new dimension to the accuracy versus time trade-off, thus expanding the landscape of possibilities in this domain.

The inherent flexibility of the architecture, coupled with the incorporation of multiple designs inspired by physics, promises to enrich future investigations in fast calorimeter simulation. Further optimization options are available, leveraging the decomposed and adaptable nature of the two-stage model. Particularly promising is the exploration of latent diffusion architecture in the second stage, offering potential for enhanced scalability and expressive design of the latent space as the simulation demands for high granularity calorimeter increase.

In conclusion, our work serves as a stepping stone towards achieving a comprehensive solution for fast and accurate calorimeter simulation, offering insights and directions for future research in this critical area of high-energy physics.

\FloatBarrier
\begin{appendix}
\numberwithin{equation}{section}

\section{Hyper-parameters}
\label{asec:arch}
The key hyperparameters of the models described in the paper are listed in Table~\ref{tab:caloVQ_parameters}
\begin{table}[]
\centering
\begin{tabular}{cccccc}
\hline
                              & ds1-photon & ds1-pion & ds2  & ds3   &ds3(normed)   \\
                              \hline
log\_transform a              & 1        & 1        & 1    & 1       & 1     \\
log\_transform b              & 8000     & 8000     & 3000 & 40000   & 40000 \\
log\_transform c              & 10       & 10       & 7    & 10      & 10   \\
\hline
Hidden layers                 & 5        & 5        & 7    & 10      & 10   \\
VQ dim                        & 256      & 256      & 256  & 192     & 256  \\
cond. dim                     & 3        & 3        & 1    & 1       & 1   \\
codebook size                 & 1024     & 1024     & 1024 & 1024    & 1024 \\
R codes/token                       & 10       & 14       & 2    & 2       & 90   \\
shower codes                  & 32       & 32       & 68   & 182     & 624  \\
stage1 \#pars/M               & 3.8      & 4.1      & 3.1  & 2.1     & 2.2 \\
\hline
xformer layer                 & 2        & 2        & 2    & 1       & 1   \\
xformer head                  & 2        & 2        & 2    & 1       & 1   \\
xformer embed                 & 64       & 64       & 64   & 16      & 128 \\
stage2 \#pars/K               & 231      & 231      & 235  & 38      & 551  \\
\hline
\end{tabular}
\caption{Main hyper-parameters and number of trainable parameters of pre-processing, first stage model and second stage model of caloVQ. For "hidden\_layers" the numbers are halved since the encoder and decoder are symmetry. }
\label{tab:caloVQ_parameters}
\end{table}

\section{Performance and Computing for Two Stages}
\label{asec:cost_s12}

The performance and computing cost for two stages are listed in Table~\ref{tab:cost0}

\begin{table}[]
\centering
\begin{tabular}{lllll}
\hline
Dataset     & N\#Pars(S1)/M & S1 time/ms & N\#Pars(S2)/M & S2 time/ms  \\
\hline
ds1-photon  & 3.83          & 0.02       & 0.23          & 0.23        \\
ds1-pion    & 4.08          & 0.02       & 0.23          & 0.26        \\
ds2         & 3.08          & 0.17       & 0.23          & 0.46        \\
ds3         & 2.12          & 0.35       & 0.04          & 0.79        \\
ds3(normed) & 2.22          & 1.70       & 0.55          & 34.40      \\
\hline
\end{tabular}
\caption{Number of parameters and generation time per shower of each stage for differrent model.}
\label{tab:cost0}
\end{table}

\end{appendix}

\clearpage
\bibliography{BiBTeX_File.bib}

\end{document}